\documentclass[12pt]{article}

\usepackage{amsmath,epsfig,ulem,cite}

\newcommand{\se}{{\tilde{e}}}

\newcommand{\sn}{{\tilde{\nu}}}

\newcommand{\mse}[1]{m_{\tilde{e}_{#1}}}

\newcommand{\mst}{m_{\tilde{t}_1}}

\newcommand{\cha}{\tilde{\chi}}
\newcommand{\neu}{\tilde{\chi}^0}
\newcommand{\mcha}[1]{m_{\tilde{\chi}^\pm_{#1}}}
\newcommand{\mneu}[1]{m_{\tilde{\chi}^0_{#1}}}
\newcommand{\sbt}{\mathswitch {s_\beta}}
\newcommand{\cbt}{\mathswitch {c_\beta}}

\newcommand{\h}[1][1]{\frac{#1}{2}}

\def\mathswitch#1{\relax\ifmmode#1\else$#1$\fi}
\def\mathswitchr#1{\relax\ifmmode{\mathrm{#1}}\else$\mathrm{#1}$\fi}
\newcommand{\PW}{\mathswitchr W}
\newcommand{\PZ}{\mathswitchr Z}
\newcommand{\PH}{\mathswitchr H}
\newcommand{\Pe}{\mathswitchr e}
\newcommand{\Pb}{\mathswitchr b}
\newcommand{\Pt}{\mathswitchr t}

\newcommand{\MW}{\mathswitch {m_\PW}}
\newcommand{\MZ}{\mathswitch {m_\PZ}}
\newcommand{\MH}{\mathswitch {m_\PH}}

\newcommand{\mb}{\mathswitch {m_\Pb}}
\newcommand{\mt}{\mathswitch {m_\Pt}}

\newcommand{\scrs}{{}}
\newcommand{\sw}{\mathswitch {s_{\scrs\PW}}}
\newcommand{\cw}{\mathswitch {c_{\scrs\PW}}}

\newcommand{\tev}{\,\, \mathrm{TeV}}
\newcommand{\gev}{\,\, \mathrm{GeV}}

\newcommand{\RR}{{\rm R}}
\newcommand{\LL}{{\rm L}}

\newcommand{\anc}{\rule{0mm}{0mm}}
\newcommand{\lesim}{\,\raisebox{-.1ex}{$_{\textstyle <}\atop^{\textstyle\sim}$}\,}
\newcommand{\gesim}{\,\raisebox{-.3ex}{$_{\textstyle >}\atop^{\textstyle\sim}$}\,}

\newcommand{\mycaption}[1]{\caption{\sl #1}}

\begin{document}
\thispagestyle{empty}

\def\thefootnote{\fnsymbol{footnote}}

\begin{flushright}
FERMILAB--PUB--05--362--E--T
\end{flushright}

\vspace{1cm}

\begin{center}

{\Large\bf Analyzing the Scalar Top Co-Annihilation Region at the ILC}
       \\[3.5em]
{\large 
M.~Carena$^{1}$,
A.~Finch$^{2}$,
A.~Freitas$^{1}$,
C.~Milst\'ene$^{1}$,
H.~Nowak$^{3}$,
A.~Sopczak$^{2}$}

\vspace*{1cm} 

{\sl
$^1$ Fermi National Accelerator Laboratory, Batavia, IL 60510-500, USA

\vspace*{0.4cm}

$^2$ Lancaster University, Lancaster LA1 4YB, United Kingdom

\vspace*{0.4cm}

$^3$ Deutsches Elektronen-Synchrotron DESY, D--15738 Zeuthen, Germany
}

\end{center}

\vspace*{2.5cm}

\begin{abstract}

The Minimal Supersymmetric Standard Model opens the possibility of electroweak
baryogenesis provided that the light scalar top quark (stop) is lighter than
the top quark. In addition, the lightest neutralino is an ideal candidate to
explain the existence of dark matter. For a light stop with mass close to the
lightest neutralino, the stop-neutralino co-annihilation mechanism becomes
efficient, thus rendering the predicted dark matter density compatible with
observations. Such a stop may however remain elusive at hadron colliders. Here
it is shown that a future linear collider provides a unique opportunity to
detect and study the light stop. The production of stops with small
stop-neutralino mass differences is studied in a detailed experimental analysis
with a  realistic detector simulation including a CCD vertex detector for
flavor tagging. Furthermore, the linear collider, by precision measurements of
superpartner masses and mixing angles, also allows to determine the dark matter
relic density with an accuracy comparable to recent astrophysical observations.

\end{abstract}

\def\thefootnote{\arabic{footnote}}
\setcounter{page}{0}
\setcounter{footnote}{0}

\newpage


\section{Introduction}

Among the most challenging questions for high energy physics research during the
next decades is the origin and stabilization of the mechanism of electroweak
symmetry breaking in particle physics and the nature of dark matter and
baryogenesis in cosmology. Both aspects
suggest the existence of new symmetries within the reach of the next generation
of colliders, thereby highlighting an interface between particle physics and
cosmology.

The existence of dark matter in the universe has been firmly established by
various experiments. Recently the dark matter abundance has been precisely
determined by the Wilkinson Microwave Anisotropy Probe (WMAP)
\cite{Spergel:2003cb}, in agreement with the Sloan Digital Sky Survey (SDSS)
\cite{Tegmark:2003ud}, $\Omega_{\rm CDM} h^2 = 0.1126^{+0.0161}_{-0.0181}$ at
the 95\% C.L. Here $\Omega_{\rm CDM}$ is the ratio of the dark matter energy
density to the critical density $\rho_{\rm c} = 3 H_0^2 / (8\pi G_{\rm N})$,
where $H_0 = h \times 100$ km/s/Mpc is the Hubble constant and $G_{\rm N}$ is
Newton's constant. One of the most attractive possibilities to explain the
nature of dark matter is the existence of weakly interacting massive particles
(WIMPs), which are stable due to an additional new symmetry.

Moreover, the generation of the baryon-anti\-baryon-asym\-metry (baryogenesis)
in the universe also demands the introduction of new physics beyond the
Standard Model. Any mechanism for baryogenesis must fulfill the three Sakharov
conditions \cite{sakharov}: (i) the existence of baryon number violating
processes, (ii) CP violation and (iii) a departure from thermal equilibrium.
Although the three conditions are fulfilled in the Standard Model, the
mechanism of electroweak baryogenesis is ruled out. Baryon number violation
arises in the Standard Model and extensions thereof in the form of
non-perturbative sphaleron processes. Out-of-equilibrium transitions can occur
at the electroweak phase transition if it is sufficiently strongly first order,
$v(T_{\rm c})/T_{\rm c} \gesim 1$, where $v(T_{\rm c})$ denotes the Higgs
vacuum expectation value at the critical temperature $T_{\rm c}$
\cite{phasetran}. A first-order phase transition can be induced by loop
contributions of light bosonic particles to the finite temperature Higgs
potential, but in the Standard Model the contribution of the  gauge bosons is
too small to allow for $v(T_{\rm c})/T_{\rm c} \gesim 1$ while obeying the
Higgs mass bound from LEP \cite{Kajantie:1996mn}. In addition, CP violation in
the Standard Model through the phase in the CKM quark mixing matrix is not
sufficient to explain the baryon asymmetry in the universe \cite{CPSM}.

One of the most compelling theories for physics beyond the Standard Model is
supersymmetry, which stabilizes the difference between the scale of electroweak
physics, $\Lambda_{\rm weak} \sim 100$ GeV, and the scale of Grand Unified
Theories (GUTs), $\Lambda_{\rm GUT} \sim 10^{16}$ GeV. The hierarchy
stabilization suggests that the masses of  the supersymmetric partners of the
Standard Model particles are near the TeV scale, within reach of the next
generation of colliders. In the presence of $R$-parity, which assigns $R=1$ to
Standard Model particles and $R=-1$ to their supersymmetric partners, the
lightest supersymmetric particle (LSP) is stable and provides a natural dark
matter candidate. 

In the Minimal Supersymmetric extension of the Standard Model (MSSM), the
superpartners of the top quark, the scalar top quarks (stops), have an
important impact on the Higgs potential \cite{CQW,EWBG,Carena:1997ki,EWBG2}.
Loop effects of light stops can induce a strongly first order electroweak phase
transition, thus generating the out-of-equilibrium condition for electroweak
baryogenesis. In addition, supersymmetric models offer new sources of CP
violation to explain the magnitude of the baryon asymmetry.

To open the window for electroweak baryogenesis, the lightest stop mass is
required to be smaller than the top quark mass, whereas the Higgs boson
involved in the electroweak phase transition must be lighter than about 120 GeV
\cite{CQW,EWBG,Carena:1997ki,EWBG2}. Hence there is an
interesting region of parameter space for which the light stop is only slightly
heavier than the neutralino LSP, thus implying that the stop-neutralino
co-annihilation process becomes significant. In the stop-neutralino
co-annihilation region consistent with the measured dark matter value, the mass
difference between the light stop and the lightest neutralino is  smaller than
about 30 GeV \cite{Balazs:2004bu}.

The Tevatron and the Large Hadron Collider (LHC) experiments will be able to
probe a light Higgs boson with Standard-Model-like couplings to the gauge
bosons, as required by electroweak baryogenesis, but the stop parameter region
compatible with dark matter is very difficult to explore at hadron colliders.
Preliminary studies for the Tevatron show that with 2--4 fb$^{-1}$ of
integrated luminosity, stops with masses up to about 170 GeV may be detected if
the stop-neutralino mass difference is larger than 30--50 GeV
\cite{Demina:1999ty}. Smaller mass differences cannot be covered due to the
reliance on a  trigger for the signature of missing transverse energy. So far,
no dedicated studies for light stop exist for the Large Hadron Collider (LHC).
However, limitations due to background levels and detector thresholds will be
even more severe at the LHC than at the Tevatron\footnote{However, at the LHC,
the production of stops from gluino decays, $pp \to \tilde{g} \tilde{g} \to t
t\, \tilde{t}^* \tilde{t}^*$, with same-sign leptonic tagging of the top
quarks, might offer additional prospect for studying stops for small mass
differences. This channel is currently under study.}.

A future international $e^+e^-$ linear collider (ILC) provides a clean
environment with relatively small background levels that allows the study of
stops for small stop-neutralino mass differences.  For instance, the Large
Electron-Positron Collider (LEP) was able to set limits on the stops even for a
mass difference to the neutralino close to 1~GeV~\cite{lep}.  Based on the
experience from LEP, the ILC might be able to explore mass differences down to
a few GeV even for stop masses comparable with the top quark mass.  In the
region of parameters where stop-neutralino co-annihilation leads to a value of
the relic density consistent with experimental results, the stop-neutralino
mass difference is never much smaller than 15~GeV, and hence an ILC seems well
suited to explore this region efficiently. 

In addition to establishing the existence of light stop quarks, the precise
measurement of their properties is crucial for testing their impact on the dark
matter relic abundance and the mechanism of electroweak baryogenesis. The
relevance of a  linear collider to probe MSSM baryogenesis through the chargino
sector have been discussed in Ref.~\cite{Murayama:2002xk}.  A linear collider
provides an excellent environment to perform precise measurements of the stop
\cite{stopsLC,stopsnew} and chargino systems~\cite{Choi:1998fh,chath}.

In this work it is shown that the ILC offers a unique possibility to elucidate
the scenario of light stop quarks and its cosmological implications. The
capabilities of the ILC for searching stops with small stop-neutralino mass
differences is investigated in detail, including realistic detector
simulations. It is studied to which accuracy the stop mass, mixing angle and
stop-neutralino mass difference can be determined at the ILC, also under the
presence of potential CP phases. This involves the analysis of the decay
spectra and production cross-sections for different beam polarizations.  From
the expected accuracies at the ILC, we will furthermore explore how precisely
the contribution of stop co-annihilation to the dark matter relic density can
be computed. 

After introducing the relevant notations and theoretical constraints and
assumptions for the light stop scenario in section~\ref{sec:not}, an
experimental study for stop searches at the linear collider is presented in
section~\ref{sec:stop}. Section~\ref{sec:param} analyzes the achievable
precision for the determination of the underlying supersymmetric parameters,
which is used in section~\ref{sec:dm} to derive the expected accuracy for the
computation of the dark matter density from linear collider measurements.
The final conclusions are given in section~\ref{sec:concl}.


\section{Notation and conventions}
\label{sec:not}

This work is restricted to the framework of the Minimal Supersymmetric
Standard Model (MSSM), including CP-violating phases in the supersymmetry
breaking terms. 

Since the Yukawa couplings of the first two generations are very small, mixing
between the L- and R-sfermions, partners of the left- and right-handed
fermions, of the first two generations can be neglected. For the supersymmetric
partners of the top quark, the scalar top quark (stop), on the other hand,
mixing effects are very important. The stop mass matrix is given by
\begin{equation}
M^2_{\tilde{t}} = \begin{pmatrix}
m_{\rm \tilde{Q}_3}^2 + \mt^2 + 
  (\tfrac{1}{2} - \tfrac{2}{3} \sw^2) \, \MZ^2 \cos 2\beta &
 \mt \bigl( A_{\rm t}^* - \mu  \cot \beta \bigr) \\
\mt \bigl( A_{\rm t} - \mu^*  \cot \beta \bigr) &
m_{\rm \tilde{U}_3}^2 + \mt^2 + \tfrac{2}{3} \,\sw^2\, \MZ^2 \cos 2\beta 
\end{pmatrix}, \label{Msf}
\end{equation}
where $\tan\beta$ is the ratio of the vacuum expectation values of the two
Higgs doublets, $m_{\rm\tilde{Q}_3}$, $m_{\rm\tilde{U}_3}$ are the left- and
right-chiral stop supersymmetry breaking masses, respectively, $A_{\rm t}$ is
the trilinear stop soft breaking parameter and $\mu$ is the Higgs/higgsino
parameter in the superpotential. Diagonalization of the stop mass matrix yields
the mass eigenvalues
\begin{equation}
m_{\tilde{t}_{1,2}}^2 = \tfrac{1}{2} \left[
  M^2_{\tilde{t},11} + M^2_{\tilde{t},22} \mp 
  \sqrt{\left(M^2_{\tilde{t},11} - M^2_{\tilde{t},22} \right)^2 + 
  4 \left|M_{\tilde{t},12}\right|^4} \right],
\end{equation}
and the mixing angle
\begin{equation}
\sin^2 \theta_{\tilde{t}} = 
  \frac{|M_{\tilde{t},12}|^4}{2 \, |M_{\tilde{t},12}|^4 
  	+ (m_{\tilde{t}_1}^2 -
	M^2_{\tilde{t},22})(M^2_{\tilde{t},11} - M^2_{\tilde{t},22})}.
\end{equation}
In this work, inter-generational (CKM) mixing effects for fermions and
sfermions are neglected.

Apart from the electroweak parameters, the spectrum of the charginos and
neutralinos is described by the Higgs/higgsino parameter $\mu$ and the soft
SU(2) and U(1) gaugino parameters, $M_2$ and $M_1$, respectively. In the
wino-higgsino basis, the chargino mass matrix reads
\begin{equation}
X = \begin{pmatrix} M_2 & \sqrt{2} \MW \sin\beta \\
                \sqrt{2} \MW \cos\beta & \mu \end{pmatrix}.
\label{eq:Mcha}
\end{equation}
The mass eigenvalues and the mixing angles for the left- and
right-chiral chargino components are obtained according to \cite{chath}
\begin{align}
\mcha{1,2}^2 &= \frac{1}{2} \left( M_2^2 + |\mu|^2 + 2 \MW^2 - \Delta_C \right),
\label{eq:cmass}\\
\cos 2\phi_{\LL,\RR} &= -(M_2^2 - |\mu|^2 \mp 2 \MW^2 \cos 2\beta) / \Delta_C.
\label{eq:cang}
\end{align}
with
\begin{equation}
\Delta_C = \sqrt{(M_2^2 - |\mu|^2)^2 + 4 \MW^4 \cos^2 2\beta +
                4 \MW^2 (M_2^2 + |\mu|^2) +
                8 \MW^2 M_2 |\mu| \sin 2\beta \cos \phi_\mu}.
\end{equation}

\noindent
The neutralino mass term in the current eigen-basis is given by
\begin{equation}
\mathcal{L}_{\rm m_{\cha^0}} = -\h {\psi^0}^\top \,
Y \, \psi^0 + \mbox{h.c.}, \qquad
\psi^0 = \bigl( \widetilde{B}, \widetilde{W}^0,
  \widetilde{H}_{\rm d}^0, \widetilde{H}_{\rm u}^0 \bigr)^{\!\top},
\label{eq:neuLagr}
\end{equation}
with the symmetric mass matrix
\begin{equation}
Y = \begin{pmatrix} M_1 & 0 & -\MZ\,\sw\,\cbt & \MZ\,\sw\,\sbt \\
                    0 & M_2 & \MZ\,\cw\,\cbt & -\MZ\,\cw\,\sbt \\
                    -\MZ\,\sw\,\cbt & \MZ\,\cw\,\cbt & 0 & -\mu \\
                    \MZ\,\sw\,\sbt & -\MZ\,\cw\,\sbt & -\mu & 0 \end{pmatrix},
\label{eq:Mneu}
\end{equation}
in which the abbreviations $\sbt = \sin\beta$ and $\cbt = \cos\beta$ have been
introduced; $\sw$ and $\cw$ are the sine and cosine of the electroweak
mixing angle. The transition to the mass eigen-basis is performed by the
unitary mixing matrix $N$,
\begin{equation}
N^*YN^{-1} = \mbox{diag}
  \bigl(\mneu{1}^2, \mneu{2}^2, \mneu{3}^2, \mneu{4}^2 \bigr).
\label{eq:neumix}
\end{equation}
Explicit analytical solutions for the mixing matrices can be found in
Ref.~\cite{chath,ckmz}\footnote{Note that the convention for the chargino mass matrix
$X$ used in Ref.~\cite{chath} is different from eq.~\eqref{eq:Mcha}.}.

In the MSSM Higgs sector, the tree-level masses of the CP-even neutral Higgs
bosons $h^0$ and $H^0$ and the charged scalar $H^\pm$ can be expressed through
the gauge boson masses, the mass of the pseudo-scalar Higgs boson, $m_{\rm
A^0}$, and $\tan\beta$. The Born relations are however significantly modified
by radiative corrections, with dominant effects originating from top and stop
loops. 

In the MSSM the Higgs mass is very sensitive to the stop spectrum. In order to
be consistent with the bound $m_{\rm h^0} \gesim 114.4 \gev$ from direct searches
at LEP \cite{lephbound} and with one light stop state, the heavier stop mass
has to be above about 1 TeV and the trilinear coupling $A_t$ has to be
sizable \cite{Carena:1997ki}.  Constraints from electroweak precision
data are satisfied when the light stop is mainly right-chiral. This is
naturally achieved for values of the stop supersymmetry breaking parameters
$m_{\rm\tilde{Q}_3}^2 \gesim 1 \tev^2$ and $m_{\rm\tilde{U}_3}^2 \lesim 0$,
respectively. The stop mixing parameter $X_t = \mu \cot \beta - A_t$ is bounded
from below by the Higgs boson mass constraint from LEP and from above by the
requirement of the strength of the first order electroweak phase transition,
leaving the allowed range $0.3 \lesim |X_t|/m_{\rm\tilde{Q}_3} \lesim 0.5$
\cite{Carena:1997ki}. The lower bound is weakened for large values of
$m_{\rm\tilde{Q}_3}$ of several TeV.

The MSSM Higgs masses with CP violation have been calculated 
including complete one-loop and leading two-loop corrections, see
{\it e.g.}  Ref.~\cite{higgsmass}. In this work, however, the process of
baryogenesis at the electroweak phase transition is computed with the program
of Refs.~\cite{Carena:2002ss,morr}, which includes only one-loop corrections to
the zero temperature Higgs potential. Since the allowed mass range for the
Higgs boson  is constrained by the mechanism of electroweak baryogenesis, for
consistency the Higgs mass is determined by the minimization of the one-loop
effective potential. This implies that only one-loop corrections are included
in the calculation of the Higgs mass as well\footnote{An analysis including
two-loop corrections to the effective potential is in progress \cite{tleff}.}.

For successful electroweak baryogenesis, an additional source of CP violation
beyond the Standard Model CKM matrix is necessary. Within the MSSM, the
dominant source are chargino loops, with a contribution proportional to
$\rm Im\{\mu M_2\}$ \cite{Carena:1997gx,Carena:2002ss}. To generate a sufficiently
large baryon asymmetry, the charginos are required to be relatively light,
$\mcha{1} \sim {\cal O}(\mbox{a few 100 GeV})$. In addition, the CP-violating
phase needs to be sizable, $\arg(\mu M_2) \gesim 0.1$ \cite{Carena:2002ss}.

A very large CP-violating phase, on the other hand, is restricted by
experimental bounds on the electric dipole moments of the electron, neutron and
$^{199}$Hg nucleus. The leading contributions from one-loop sfermion-gaugino
loops \cite{edm1,edm2} become small for large masses of the first two
generation sfermions of several TeV. Furthermore, CP-violating phases in the
sfermion $A$-parameters can result in cancellations for the electric dipole
moments \cite{edm1}, without having much effect on electroweak baryogenesis.
For the present work, the strongest constraint arises from the bound on the
electric dipole moment of the electron, $|d_\Pe| < 1.6 \times 10^{-27} \, e$~cm
\cite{Regan:2002ta}, effectively restricting the allowed MSSM parameter space.

A CP-violating phase in $M_2$ can always be transferred into the $\mu$ parameter
by means of a unitary transformation. In principle, there can also be
non-trivial phases in the gaugino parameters $M_1$ and $M_3$, but their effect
on electroweak baryogenesis is small. Therefore in the following all gaugino
soft parameters are assumed real, while the generation of the baryon asymmetry
is connected with a phase in the $\mu$ parameter,
\begin{equation}
\mu = |\mu| \times e^{i \phi_\mu}.
\end{equation}
For simplicity, this report will only investigate scenarios where the lightest
neutralino is bino-like, {\it i.e.} $M_1 \ll M_2,|\mu|$ and the annihilation
cross-section is importantly enhanced through co-annihilation with the stop.
For small values of $\mu$, the LSP can acquire a significant Higgsino
component, which increases the annihilation via s-channel $Z$ and Higgs boson
exchange. This case will be considered elsewhere.


\section{Analysis of light stops at a linear collider}
\label{sec:stop}

For small mass differences $\Delta m = \mst - \mneu{1}$, the dominant decay
mode of the light stop quark, $\mst < \mt$, is into a charm quark and the
lightest neutralino, $\tilde{t}_1 \to c \, \neu_1$. In the parameter region
where the co-annihilation mechanism becomes efficient, the typical
mass differences are $\Delta m < 30$ GeV. Thus the two- and three-body decay
modes $\tilde{t}_1 \to t \, \neu_1$, $\tilde{t}_1 \to b \, \cha^+_1$,
$\tilde{t}_1 \to b \, W^+ \, \neu_1$ and $\tilde{t}_1 \to b \, l^+ \sn_l$ are
kinematically forbidden, since the sneutrino and chargino are assumed to be
heavier than the light stop. Since the decay $\tilde{t}_1 \to c \, \neu_1$
occurs only at one-loop order, it is formally of the same order ${\cal
O}(\alpha^3)$ as the four-body decay $\tilde{t}_1 \to b \, l^+ \, \nu_l \,
\neu_1$. However, the decay into a charm quark is sensitive to the
supersymmetric flavor structure at the supersymmetry breaking scale
\cite{Hikasa:1987db}. 
For models with high-scale supersymmetry breaking, this therefore introduces a
large enhancement factor for this process.
In the following it is assumed that the 
branching ratio for $\tilde{t}_1 \to c \, \neu_1$ is 100\%.

In this section the production of light stops at a $\sqrt{s} = 500$ GeV linear
collider is analyzed. The analysis makes use of high luminosity ${\cal L} \sim
500 {\rm \ fb}^{-1}$ and polarization of both beams. At an $e^+e^-$ collider,
scalar top quarks could be produced in pairs,
\begin{equation}
e^+e^- \to \tilde{t}_1 \, \tilde{t}_1^* \to c \neu_1 \, \bar{c} \neu_1,
\end{equation}
leading to a signature of two charm jets plus missing energy. The tree-level
cross-section reads
\begin{equation}
\sigma[e^+e^- \to \tilde{t}_1 \tilde{t}_1^*] =
  \frac{\pi\alpha^2}{2s} \left(1 - 4 \mst^2/s\right)^{3/2} \;
  \sum_{i=\LL,\RR} P_i
  \left[ \frac{2}{3} + g_i \, \frac{3 \cos^2 \theta_{\tilde{t}} - 
  4\sw^2}{6\sw\cw} \,
  \frac{s}{s-\MZ^2}  \right]^2, \label{eq:xsec}
\end{equation}
with
\begin{equation}
g_\LL = \frac{-1+2\sw^2}{2\sw\cw}, \qquad
g_\RR = \frac{\sw}{\cw}, \qquad
P_{\LL,\RR} = \big[1\mp P(e^-)\big]\big[1\pm P(e^+)\big],
\end{equation}
where $P(e^\pm)$ are the polarization degrees of the $e^\pm$ beams and
negative/positive values indicate left-/right-handed polarization.

For small $\Delta
m$, the jets are relatively soft and it is challenging to separate them from
Standard Model backgrounds. Since small differences in the expected signal and
background distributions contribute to the selection of the signal events,
a realistic detector simulation is applied.
Both the signal and background events are generated with {\sc Pythia 6.129}
\cite{pythia}, which has been adapted with a private code for the stop signal 
generation \cite{stopgen}
and preselection previously used in Ref.~\cite{stopsnew}. 
The {\sc Simdet} detector simulation \cite{simdet} 
has been used, describing a typical ILC detector.
The analysis is based on the tool
{\sc N-Tuple} \cite{ntuple}, which incorporates jet finding
algorithms.  The cross-sections of the signal process and the relevant
background processes have been computed by adapting the Monte-Carlo code used
in Ref.~\cite{slep} and by {\sc Grace 2.0} \cite{grace}, with cross-checks with
{\sc CompHep 4.4} \cite{comphep} for most processes. Table~\ref{tab:xsec}
summarizes the expected signal and background cross-sections. To
avoid the infrared divergence of the two-photon background, it is given with a
cut on the minimal transverse momentum, $p_{\rm t} > 5$ GeV. 

All processes are simulated including beamstrahlung for cold ILC technology
as parameterized in the program {\sc Circe 1.0} \cite{circe}. The effect of
beamstrahlung is twofold. Firstly, due to electromagnetic interactions between the
two densely charged incoming bunches, a spread of the $e^+/e^-$ beam energy
spectra is generated. On the other hand, these electromagnetic interactions
result in the radiation of (typically soft) photons that can interact and lead
to two-photon events in addition to the normal two-photon background mentioned
above. In Table~\ref{tab:xsec}, the combined two-photon background from all
sources is listed.

As evident from the table, the Standard Model backgrounds are several orders of
magnitude larger than the signal and need to be reduced with efficient
selection cuts. Backgrounds from supersymmetric processes are typically small
\cite{stopsnew}. If the stop is very light and the light chargino can decay
into the stop and a bottom quark, $\cha^+_1 \to \tilde{t}_1 \, \bar{b}$, the
largest supersymmetric background comes from chargino pair production. Although
this process has a four-jet signature with two b-quark jets and two c-quark
jets, a fraction of events with overlapping jets can  be a relevant background
to stop pair production, since the chargino production cross-section is
typically large, of the order of 1 pb.\footnote{Depending on the supersymmetry
scenario, chargino pair production can actually be an interesting discovery
process for light stops. However, this possibility will not be explored further
in this paper.} The contribution of this background will be addressed later in
this section.

\renewcommand{\arraystretch}{1.2}%
\begin{table}[tb]
\centering
\begin{tabular}{|l|rrr|}
\hline
Process &  \multicolumn{3}{c|}{Cross-section [pb]} \\
\hline
$P(e^-) / P(e^+)$ & 0/0 & $-$80\%/+60\% & +80\%/$-$60\% \\
\hline
$\tilde{t}_1 \tilde{t}_1^* \quad$ $\mst = 120 \gev$ & 0.115 & 0.153 & 0.187 \\
\phantom{$\tilde{t}_1 \tilde{t}_1^* \quad$} $m_{\tilde{t}_1} = 140$ GeV &
  0.093 & 0.124 & 0.151 \\
\phantom{$\tilde{t}_1 \tilde{t}_1^* \quad$} $m_{\tilde{t}_1} = 180$ GeV &
  0.049 & 0.065 & 0.079 \\
\phantom{$\tilde{t}_1 \tilde{t}_1^* \quad$} $m_{\tilde{t}_1} = 220$ GeV &
  0.015 & 0.021 & 0.026 \\
\hline
$W^+W^-$ & 8.55 & 24.54 & 0.77 \\
$ZZ$	& 0.49 & 1.02 & 0.44 \\
$W e\nu$ & 6.14 & 10.57 & 1.82 \\
$e e Z$  & 7.51 & 8.49 & 6.23 \\
$q \bar{q}$, $q \neq t$ & 13.14 & 25.35 & 14.85 \\
$t \bar{t}$ & 0.55 & 1.13 & 0.50 \\
2-photon, $p_{\rm t} > 5$ GeV & 936\phantom{.00}&& \\
\hline
\end{tabular}
\mycaption{Cross-sections for the stop signal and Standard Model background
processes for $\sqrt{s} = 500 \gev$ and different polarization combinations. 
The signal is given for $\cos \theta_{\tilde{t}} = 0.5$ and
different values of the stop mass. Negative polarization values refer to
left-handed polarization and positive values to right-handed polarization.}
\label{tab:xsec}
\end{table}
\renewcommand{\arraystretch}{1}%

In the first step of the event selection, the following preselection
cuts are applied:
\begin{equation}
\begin{aligned}
&4 < N_{\mbox{\scriptsize charged tracks}} < 50, &\quad &p_{\rm t} > 5 \gev, \\
&|\cos \theta_{\rm thrust}| < 0.8, & & |p_{\rm long,tot}/p_{\rm tot}| < 0.9, \\
&E_{\rm vis} < 0.75 \sqrt{s}, & & m_{\rm inv} < 200 \gev.
\end{aligned}
\end{equation}
The lower cut on the number of charged tracks removes most of the background
with leptons and no jets and the upper cut part of the $t\bar{t}$ background. 
By requiring a minimal transverse momentum $p_{\rm t}$, the two-photon
background and back-to-back processes like $q\bar{q}$ are largely reduced.
Since for most background processes a potential momentum or energy imbalance 
occurs from particles lost in the beam pipe cone, a cut on the thrust angle
$\theta_{\rm thrust}$ and the longitudinal momentum $p_{\rm long,tot}$ reduces
all backgrounds. The signal, on the other hand, has very little longitudinal
thrust since it follows a polar $\sin^2 \theta$ distribution. Furthermore, the
signal is characterized by a large amount of missing energy,  corresponding to
visible energy $E_{\rm vis}$ substantially smaller than the center-of-mass
energy. The requirement of missing energy mostly reduces processes without
neutrinos in the final state, {\it i.e.} $q\bar{q}$ and hadronic decays of
$W^+W^-$, $ZZ$, $t\bar{t}$. Finally, demanding that the total visible invariant
mass is below 200 GeV decreases $W^+W^-$, $q\bar{q}$ and $t\bar{t}$ backgrounds
without affecting the signal, which typically leads to relatively soft events
for small mass differences $\Delta m$.

\renewcommand{\arraystretch}{1.18}%
\begin{table}[tb]
\centering
\begin{tabular}{|l|r|r|rrrrrr|r|}
\hline
 &  & After$\;$\anc &  & &  &  &  & 	& Scaled to \\[-.5ex]
Process & Total & presel. & cut 1 & cut 2 & cut 3 & cut 4 & cut 5 & cut 6
	& 500 fb$^{-1}$ \\
\hline
$W^+W^-$ & 210,000 & 2814 & 827 & 28 & 25 & 14 & 14 & 8 & 145 \\ 
$ZZ$	& 30,000 & 2681 & 1987 & 170 & 154 & 108 & 108 & 35 & 257 \\ 
$W e\nu$ & 210,000 & 53314 & 38616 & 4548 & 3787 & 1763 & 1743 & 345 & 5044 \\ 
$e e Z$  & 210,000 & 51 & 24 & 20 & 11 & 6 & 3 & 2 & 36 \\ 
$q \bar{q}$, $q \neq t$ & 350,000 & 341 & 51 & 32 & 19 & 13 & 10 & 8 & 160 \\ 
$t \bar{t}$ & 180,000 & 2163 & 72 & 40 & 32 & 26 & 26 & 25 & 38 \\
2-photon & $8 \times 10^6$ & 4061 & 3125 & 3096 & 533 & 402 & 0 & 0 & $<164$\\
\hline
$\mst = 140$ &&&&&&&&&\\
\anc $\; \Delta m = 20$ & 50,000 & 68.5 & 48.8 & 42.1 & 33.4 &
	27.9 & 27.3 & 20.9 & 9720 \\ 
\anc $\; \Delta m = 40$ & 50,000 & 71.8 & 47.0 & 40.2 & 30.3 &
	24.5 & 24.4 & 10.1 & 4700 \\ 
\anc $\; \Delta m = 80$ & 50,000 & 51.8 & 34.0 & 23.6 & 20.1 &
	16.4 & 16.4 & 10.4 & 4840 \\ 
$\mst = 180$ &&&&&&&&&\\
\anc $\; \Delta m = 20$ & 25,000 & 68.0 & 51.4 & 49.4 & 42.4 &
	36.5 & 34.9 & 28.4 & 6960 \\ 
\anc $\; \Delta m = 40$ & 25,000 & 72.7 & 50.7 & 42.4 & 35.5 &
	28.5 & 28.4 & 20.1 & 4925 \\ 
\anc $\; \Delta m = 80$ & 25,000 & 63.3 & 43.0 & 33.4 & 29.6 &
	23.9 & 23.9 & 15.0 & 3675 \\ 
$\mst = 220$ &&&&&&&&&\\
\anc $\; \Delta m = 20$ & 10,000 & 66.2 & 53.5 & 53.5 & 48.5 &
	42.8 & 39.9 & 34.6 & 2600 \\ 
\anc $\; \Delta m = 40$ & 10,000 & 72.5 & 55.3 & 47.0 & 42.9 &
	34.3 & 34.2 & 24.2 & 1815 \\ 
\anc $\; \Delta m = 80$ & 10,000 & 73.1 & 51.6 & 42.7 & 37.9 &
	30.3 & 30.3 & 18.8 & 1410 \\ 
\hline
\end{tabular}
\mycaption{Total number of events generated for background and signal 
as well as background
event numbers and $\tilde{t}_1 \tilde{t}_1^*$  signal efficiencies in \% (for
some examples of  $\mst$ and $\Delta m$ in GeV) after preselection and each of
the final selection cuts. In the last column the expected event numbers are
scaled to a luminosity of 500~fb$^{-1}$, for zero beam polarization. The cuts
are explained in the text.}
\label{tab:evtn}
\end{table}
\renewcommand{\arraystretch}{1}%
Table~\ref{tab:evtn} shows the number of events generated for each background
process and the number of events left after these preselection cuts. Also shown
is the efficiency of the signal after the preselection for various values of the stop and neutralino
masses. About 70\% of the signal events pass the preselection for most of the
investigated stop masses and mass differences $\Delta m$.

For the final event selection the following cuts are applied (see
Table~\ref{tab:evtn}):
\begin{enumerate}

\item The number of reconstructed jets is required to be exactly two. Jets are
reconstructed with the Durham algorithm with the jet resolution parameter
$y_{\rm cut} = 0.003 \times \sqrt{s}/E_{\rm vis}$. As noted in
Ref.~\cite{opal}, the $E_{\rm vis}$-dependent $y_{\rm cut}$ parameter is
crucial for effective di-jet reconstruction over a wide range of stop and
neutralino parameters. The coefficient 0.003 was tuned to most effectively
reject four-jet $W^+W^-$ events while preserving most of the signal. The cut
reduces substantially the number of $W$ and quark pair events.

\item The limit on the visible energy $E_{\rm vis}$ in the preselection is
lowered to $E_{\rm vis} < 0.4 \sqrt{s}$ in order to cut down on $W^+W^-$, $ZZ$
and di-quark events. In addition, a window for the invariant jet mass around the
$W$-boson mass, 
$70 \gev < m_{\rm jet,inv} < 90 \gev$,
is excluded to reduce the large $W e \nu$ background.

\item The acollinearity $\phi_{\rm acol}$ is defined as the opening angle between
the two jets. Events from $e^+e^- \to q\bar{q}$ and $\gamma\gamma \to q\bar{q}$
processes with back-to-back topology are removed by requiring  $\cos \phi_{\rm
acol} > -0.9$. 

\item By applying a more severe cut on the thrust angle than in the
preselection, $|\cos \theta_{\rm thrust}| < 0.7$, events with $W$ bosons are
further reduced.

\item The remaining two-photon background is almost completely removed by
increasing the cut on the transverse momentum, $p_{\rm t} > 12 \gev$.

\item At this point, the largest remaining background is from single $W$
production, $e^+e^- \to W e\nu$. It resembles the signal closely in most
distributions, {\it e.g.} as a function of the visible energy, thrust or
acollinearity. 
The only distinctive kinematical discrimination is through the invariant mass
distribution, which has a resonance around the $W$-boson mass. 
In addition, the use of charm tagging helps to improve the signal-to-background
ratio. Here, the charm tagging is implemented using a neural network analysis as
described in Ref.~\cite{kuhlhiggs}. The neural network has been optimized to
single out the desired two-charm events by training it to 
efficiently reduce the $We\nu$ background while preserving the stop signal
for small mass differences. 
For each event, it yields a probability that this event contains two relatively
soft charm jets. Only events with an identification probability of more than 40\%
are kept.
Since the charm tagging alone is not sufficient to fully reduce the single-$W$
background, the excluded invariant jet mass window from
cut 2 is increased to 
($60 \gev < m_{\rm jet,inv} < 90 \gev$), 
at the cost of losing a substantial amount of the signal for
certain values of the stop and neutralino masses.

\end{enumerate}

All cuts have been optimized to preserve  signal events for small mass
differences. The resulting numbers of events, scaled to a luminosity of
500~fb$^{-1}$, and the signal efficiencies are summarized in
Table~\ref{tab:evtn}. After the final selection, between 10\% and 35\% of the
signal is remaining. The expected $\tilde{t}_1 \tilde{t}_1^*$ signal event
numbers are of same order of magnitude as for the Standard Model background, $N
\sim {\cal O}(10^4)$.

As mentioned above, if the stop is lighter than the light chargino, a
potentially large background can originate from chargino pair production, with
the subsequent decay $\cha^+_1 \to \tilde{t}_1 \, \bar{b}$. Since this
background is characterized by four partons in the final state including two
bottom quarks, it is largely removed by the selection of two-jet events and the
charm tagging. After applying all cuts listed above, the chargino background is
reduced by a factor of about 5000, corresponding to about 100 remaining events
for 500~fb$^{-1}$ luminosity. It will therefore be neglected in the following.

In order to explore the reach of a future linear collider for very small mass
differences $\Delta m = \mst - \mneu{1}$, signal event samples have been
generated also for  $\Delta m = 10 \gev$ and 5 GeV. The results are listed in
Table~\ref{tab:sigeff}.
\renewcommand{\arraystretch}{1.2}%
\begin{table}[tb]
\centering
\begin{tabular}{|r|r@{}cccc|}
\hline
$\Delta m$ & $m_{\tilde{t}_1} =\;$ & 120 GeV & 140 GeV & 180 GeV & 220 GeV \\
\hline
80 GeV && & 10\% & 15\% & 19\% \\
40 GeV && & 10\% & 20\% & 24\% \\
20 GeV && 17\% & 21\% & 28\% & 35\% \\
10 GeV && 19\% & 20\% & 19\% & 35\% \\
5 GeV &&  2.5\% & 1.1\% & 0.3\% & 0.1\% \\
\hline
\end{tabular}
\mycaption{Signal efficiencies for $\tilde{t}_1 \tilde{t}_1^*$ production after
final event selection for different combinations of the stop mass $\mst$ and
mass difference $\Delta m = \mst - \mneu{1}$.
}
\label{tab:sigeff}
\end{table}
\renewcommand{\arraystretch}{1}%
As evident from the table, the efficiency drastically deteriorates for $\Delta m
= 5 \gev$, as a result of the $p_{\rm t}$ cut (cut 5). An optimization of the
event selection for very small mass differences will be addressed in a future
study.

\begin{figure}[tb]
\centering
\epsfig{file=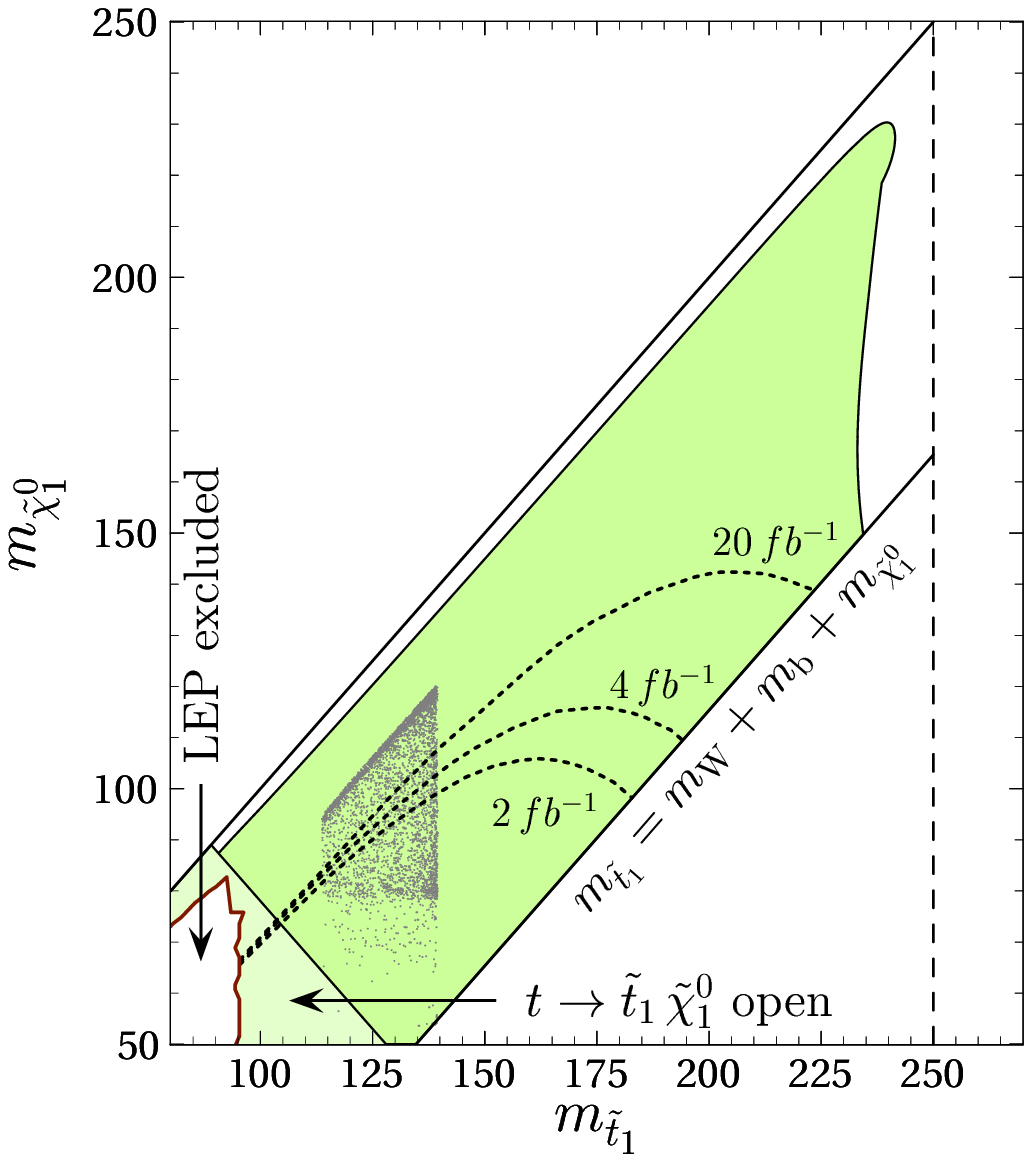, height=10cm, bb=15 340 315 682}
\epsfig{file=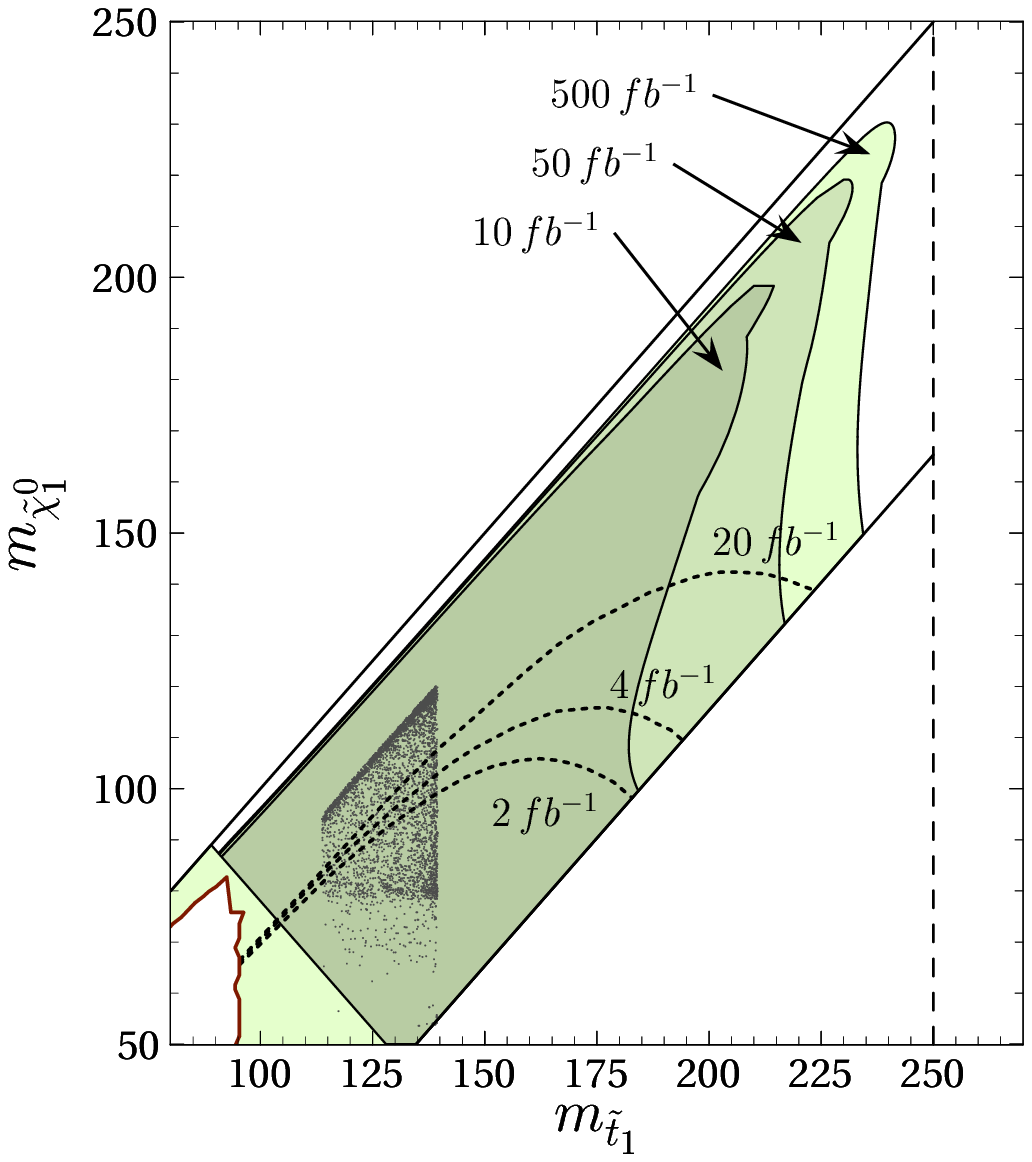, height=10cm, bb=59 340 305 682, clip=true}
\vspace{-1.5ex}
\mycaption{\textbf{Left:} Discovery reach of linear collider with 500 fb$^{-1}$ 
luminosity and unpolarized beams at
$\sqrt{s} = 500$ GeV for production of light stops, $e^+e^- \to
\tilde{t}_1 \, \tilde{t}_1^* \to c \neu_1 \,\bar{c} \neu_1$. The results
are given in the stop vs. neutralino mass plane. In the dark shaded region, a
5$\sigma$ discovery is possible. The region where $\mneu{1} > \mst$ is
inconsistent with a neutralino LSP, while for $\mst > \MW + \mb + \neu_1$ the
three-body decay $\tilde{t}_1 \to W^+ \bar{b} \neu_1$ becomes accessible and
dominant.  In the light shaded corner to the lower left, the decay of the top
quark into a light stop and neutralino is open. The dark gray dots indicate the
region consistent with baryogenesis and dark matter.
Also shown are the parameter
region excluded by LEP searches \cite{lep} (white area in lower left) and
the projected Tevatron light stop reach \cite{Demina:1999ty} (dotted lines) for 
various integrated luminosities.
\textbf{Right:} The same for different linear collider luminosities.
}
\label{fig:cov}
\end{figure}

Based on the results for residual background levels and signal efficiencies
from the experimental simulations, the discovery reach of a 500 GeV $e^+e^-$
collider can be estimated. Figure~\ref{fig:cov} shows the discovery reach in
the parameter plane of $\mst$ and $\mneu1$. The signal efficiencies for the
parameter points Table~\ref{tab:sigeff} are interpolated to cover the whole
parameter region of interest. Then, the signal rates are computed from the
production cross-section $\sigma$ for each combination
$(\mst,\mneu1)$, multiplied by the efficiency $\epsilon$ obtained from
the simulations and the luminosity ${\cal L}$, resulting in the expected signal
event number $S = \epsilon {\cal L} \sigma$. Together with the number of
background events $B$, this yields the significance $S/\sqrt{S+B}$. The dark
shaded (green) area in  the figure corresponds to the region where a discovery
with five standard  deviations is possible, $S/\sqrt{S+B} > 5$.

In Fig.~\ref{fig:cov}, the region consistent with baryogenesis and the dark
matter abundance measured by WMAP is indicated by the dark gray points. Here
the higher density of points at the upper edge of this region corresponds to
the area where stop-neutralino co-annihilation is effective, extending to mass
differences $\Delta m$ of about 15~GeV. A linear collider could find light stop
quarks for mass differences down to $\Delta m \sim {\cal O}(5 \gev)$, covering
the complete co-annihilation region. The right plot in Fig.~\ref{fig:cov} shows
that even a relatively small integrated luminosity of 10 fb$^{-1}$ is
sufficient for this purpose. The results in Fig.~\ref{fig:cov} are given for
unpolarized beams, but remain about the same if half of the luminosity is spent
for left/right and right/left-polarized $e^+$/$e^-$ beams, respectively. For an
integrated luminosity of 500 fb$^{-1}$,  the reach extends to stop masses
$\mst$ up to almost the beam energy $\sqrt{s}/2 = 250$ GeV. In the plot, there
is a dip in the covered region at  roughly $\mst=240$ GeV and $\mneu{1}=170$
GeV.  This is a result of the cut for  the invariant di-jet mass around the $W$
boson mass (cut 6).


\section{Parameter determination}
\label{sec:param}

The discovery of light scalar top quarks, in conjunction with a
Standard-Model-like Higgs boson with a mass near 120 GeV,  would be 
a strong indication that electroweak baryogenesis is
the mechanism for the generation of the baryon asymmetry.
At the same time, supersymmetry could also 
explain the existence of dark matter in the universe, based
on the co-annihilation mechanism. In order to confirm this exciting picture,
the relevant supersymmetry parameters have to be measured accurately.

One needs to (i) determine that the light stop is mainly right-chiral to
contribute appropriately to the electroweak phase transition while being in
agreement with electroweak precision measurements, (ii) check that the masses
and compositions of the gauge/Higgs superfield sector are compatible with the
values required for the generation of the  baryon asymmetry, and (iii) compute
the dark matter annihilation cross-sections and the relic abundance so as to
compare with cosmological observations. If stop-neutralino co-annihilation is
relevant it is important to determine the stop-neutralino mass difference very
precisely. 

In this section, the experimental determination of
the relevant parameters in the stop and neutralino/chargino sectors will be
discussed. As a first step, the analysis is based on tree-level formulae. In
general, however, radiative corrections can be important and introduce a
dependence on parameters of other sectors of the supersymmetric theory. This
will be studied in a forthcoming publication.

In the following, a specific MSSM parameter point will be considered, as defined
in the appendix.
At tree-level the resulting neutralino and stop masses and mixing angle are:
\begin{equation}
\begin{aligned}
m_{\tilde{t}_1} &= 122.5  \gev, & 
m_{\tilde{t}_2} &= 4203 \gev,	& 
\cos \theta_{\tilde{t}} &= 0.0105, & \qquad
\mneu{1} &= 107.2 \gev.
\end{aligned}
\end{equation}
Note that the stop mixing angle in this scenario is very small, resulting in an
almost completely right-chiral light stop state.
The mass difference $\Delta m = m_{\tilde{t}_1} - \mneu{1} = 15.2 \gev$ lies
well within the sensitivity range of the linear collider.


\subsection{Stop quark parameters}

From the measurement of $\tilde{t}_1 \tilde{t}_1^*$ production cross-section
for different beam polarization combinations, both the mass of the light stop
and the stop mixing angle can be extracted \cite{bartl97}.  Here is it assumed
that 250 fb$^{-1}$ are spent for $P(e^-)$/$P(e^+)$ = $-$80\%/+60\% and
+80\%/$-$60\%, respectively, where negative polarization degrees indicate
left-handed polarization and positive values correspond to right-handed
polarization. Typical values for the stop production cross-sections for these
beam polarizations are given in Table~\ref{tab:xsec}. The remaining background
is summarized in Table~\ref{tab:bkgd}.
\renewcommand{\arraystretch}{1.2}%
\begin{table}[tb]
\centering
\begin{tabular}{|r|ccc|}
\hline
$P(e^-) / P(e^+)$ & 0/0 & $-$80\%/+60\% & +80\%/$-$60\% \\
\hline
Residual background cross-section [fb] & 11.7 & 20.4 & 4.3 \\
\hline
\end{tabular}
\mycaption{Residual background level after final event selection for different
beam polarizations.}
\label{tab:bkgd}
\end{table}
\renewcommand{\arraystretch}{1}%
With these numbers, the cross-sections and statistical errors
in the scenario defined in the appendix are
$\sigma[-80\%$/$+60\%] = (72 \pm 1.6)$ fb and
$\sigma[+80\%$/$-60\%] = (276 \pm 2.2)$ fb.\footnote{The cross-section values
given here differ from Table~\ref{tab:xsec} due to the different
stop mixing angle.}
Besides the statistical error, the following systematic errors for the
cross-section measurement are included: 
\begin{itemize}

\item
The mass of the lightest neutralino can be measured very precisely from the
electron decay energy spectrum of the R-selectron, if the R-selectron is light
enough to be produced in pairs at the linear collider. As described in detail in
the next subsection, a precision of $\delta
\mneu{1} = 0.1 \gev$ can be achieved with this method. The impact on the
cross-section measurements in about 0.1\%.

\item
It is assumed that the beam polarization degree can be determined with an
uncertainty of $\delta P(e^\pm)/P(e^\pm) = 0.5\%$, which is a conservative
estimate \cite{power}. The effect of the polarization uncertainty on the
stop cross-section determination is below 0.01\%.

\item
In order to extract total cross-sections, the delivered luminosity needs to be
determined, which can be measured quite accurately. According to
Ref.~\cite{lumierr}, a precision of  $\delta {\cal L} / {\cal L} = 2 \times
10^{-4}$ seems feasible, but for this work a more conservative value of $\delta
{\cal L} / {\cal L} = 5 \times 10^{-4}$ is assumed.

\item
For the theoretical simulation of the Standard Model background $B$, a relative
error of $\delta B / B = 0.3\%$ is assigned.
  This estimate is based on the $W e \nu$ process as the largest background.
  While a complete NLO calculation of that process is still missing, a 
  recent result for the related process of $W$ pair production \cite{wwnlo}  
  suggests that a NLO calculation of $W e \nu$ is feasible within the next 
  years with a remaining error well below 0.5\%.

\item 
Since the decay $\tilde{t}_1 \to c \neu_1$ is loop-suppressed, the
expected decay width is below 1 keV and the stop will hadronize before decay.
The formation of the stop hadron and the fragmentation function of stop
production cannot be predicted reliably today, leading to large errors in the
stop searches at LEP \cite{lep}. However, if the stop is discovered at the
linear collider, the uncertainties in the hadronization and fragmentation
models can be reduced substantially by using experimental data
\cite{stopfrag}. Therefore the impact of this error on the cross-section
determination is estimated to be at most 1\%.

\item 
The jet analysis and the charm tagging depends on the charm fragmentation
function. Information about charm tagging can be gained at the linear collider
itself from large samples of charm jets from Standard Model processes, such as
$Z \to c\bar{c}$. Thus the remaining error should be small, below 0.5\%.

\item
The realistic simulation of detector effects on the selection efficiency is
limited by Monte-Carlo statistics and detector calibration.
With the progress of future computing resources, the error stemming from
Monte-Carlo statistics should be negligible. On the other hand,
experience from LEP2 shows that the calibration uncertainty can be kept below
0.5\% and should further improve with future detector research and development.

\item
For the computation of the cross-sections, beamstrahlung effects need to be
understood precisely. The beamstrahlung spectrum can be extracted from Bhabha
scattering with good accuracy \cite{beammoenig}. The resulting error on the stop
cross-section is found to be about 0.02\%.

\item
Finally, higher-order radiative corrections to the stop cross-section
need to be under control. While no radiative
corrections are included in this study, it is assumed that the relevant
radiative corrections will be available at the time of start-up of the ILC,
leaving a negligible theoretical error.

\end{itemize}
Adding up the individual error source in quadrature leads to a total systematic
error of 1.3\% for $P(e^-)$/$P(e^+) = -$80\%/+60\% polarization and 1.2\% for
+80\%/$-$60\% polarization, respectively. However, the above error estimate is
rather conservative, giving the maximum expected error. Experience from many
analyses at LEP and other colliders implies that data-driven systematics can be
reduced to the level of the statistical error. This would in particular apply to
the error from hadronization, fragmentation and detector calibration. Since all
other systematic error sources are small, the total systematic error under this
assumption is about the same as the statistical error of the collected signal
events, {\it i.e.} about 0.8\%.
Combining statistical and systematic uncertainies, the total error for the
cross-section determination amounts to 1.1\% for $P(e^-)$/$P(e^+) = -$80\%/+60\%
polarization and 2.4\% for +80\%/$-$60\% polarization, respectively.

\begin{figure}[tb]
\centering
\epsfig{file=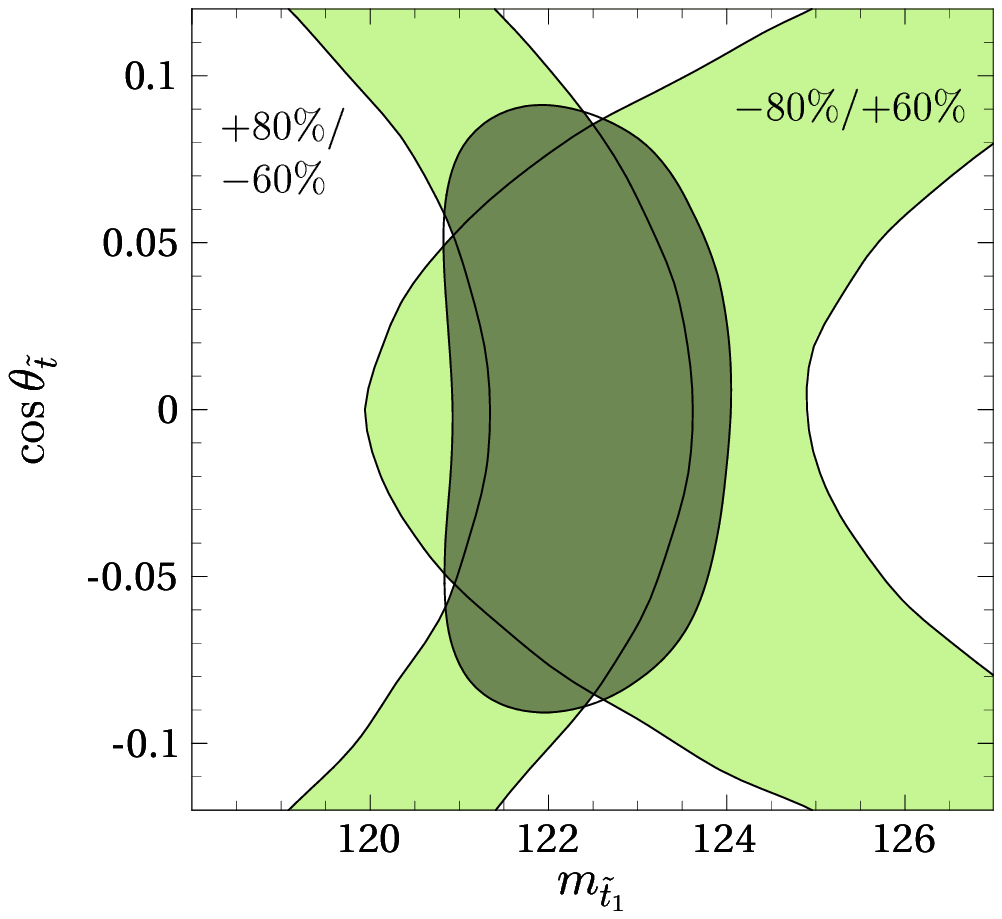, width=9cm, bb=20 422 306 682}
\mycaption{Determination of light stop mass $\mst$ and stop mixing angle
$\theta_{\tilde{t}}$ from measurements of the cross-section $\sigma(e^+e^- \to
\tilde{t}_1 \tilde{t}_1^*)$ for beam polarizations $P(e^-)$/$P(e^+)$ = 
$-$80\%/+60\% and
+80\%/$-$60\%. The widths of the light shaded bands indicate the 1$\sigma$ errors
of the cross-section measurements, which are combined into the 
1$\sigma$ two-parameter allowed region (dark shaded).
The plot includes statistical and systematic errors.}
\label{fig:stoppar}
\end{figure}
Each of the two cross-section measurements for $P(e^-)$/$P(e^+)$ =
$-$80\%/+60\% and +80\%/\linebreak[0]$-$60\% results in a band in the parameter
plane of the stop mass and mixing angle, see Fig.~\ref{fig:stoppar}. The widths
of the bands reflects the uncertainty from statistical and systematic errors.
Combining the
two cross-section measurements, the resulting precision for the stop parameter
extraction is
\begin{equation}
m_{\tilde{t}_1} = (122.5 \pm 1.0) \gev, \qquad
|\cos \theta_{\tilde{t}}| < 0.074 \quad \Rightarrow
|\sin \theta_{\tilde{t}}| > 0.9972.
\end{equation}
As the stop mixing angle in this scenario is very small,
$\cos \theta_{\tilde{t}} = 0.0105$, it cannot be experimentally distinguished
from the zero mixing case. However, a strong upper bound on the mixing angle can
be derived from the measurement.


\subsection{Chargino and neutralino parameters}

In Ref.~\cite{chath, chaex}, a general strategy for extracting the chargino and
neutralino parameters  is described.  It relies on cross-section measurements
of chargino and neutralino pair production at the linear collider and mass
chargino and neutralino mass measurements from the linear collider and LHC. The
mass determination of neutralinos and charginos at the LHC makes use of
kinematical edges in the invariant mass spectra of the decay products of squark
chains \cite{lhcsusymass}. However, in the scenario under study here most of
the squark states are assumed to be heavy to avoid the present electric dipole
moment constraints \cite{edm2}. The only light state is the light stop, which
is predominantly right-chiral and directly decays to the bino LSP. As
a consequence, the rate of chargino and neutralino production from squark
cascades in this scenario is very small. In principle, charginos and
neutralinos can also be produced directly through s-channel $Z$-boson and
photon exchange, but the rates for these processes are also relatively small
\cite{tadas}, rendering a reliable mass measurement impossible.

This analysis therefore has to be restricted to linear collider measurements.
The following observables are
included for extracting the fundamental gaugino/higgsino MSSM parameters:
\begin{itemize}
\item Chargino and neutralino mass measurements.
\item Cross-section for pair production of the light chargino.
\item Production cross-sections for the lightest and next-to-lightest
neutralinos.
\end{itemize}
This study is restricted to the framework of the MSSM, assuming two chargino and
four neutralino states with the free parameters $M_1$, $M_2$, $\mu$ and
$\tan\beta$.
For the sample scenario in the appendix, the chargino and neutralino masses at
tree-level amount to
\begin{equation}
\begin{aligned}
\mneu{1} &= 107.2 \gev, & \mneu{2} &= 196.1 \gev, & \mcha{1} &= 194.3 \gev, \\
\mneu{3} &= 325.0 \gev,	& \mneu{4} &= 359.3 \gev, & \mcha{2} &= 358.1 \gev. \\
\end{aligned}
\end{equation}
Furthermore, the scenario has been chosen such that the lighter, predominantly
right-chiral, selectron state is accessible at the linear collider,
$m_{\tilde{e}_1} = 204.2 \gev$, while the other selectron and sneutrino states
are heavy, $m_{\tilde{e}_1},m_{\tilde{\nu}_e} = 2 \tev$.

In this case,
the mass of the lightest neutralino can be determined very precisely from the
electron energy spectrum in selectron decay, $\se_1^\pm \to e^\pm \neu_1$.
Since the selectron is a scalar particle, the decay electron energy is to first
approximation  uniformly distributed between the minimum and maximum value
$E_{\rm min}$ and $E_{\rm max}$, with the distribution edges being related to
the selectron and neutralino masses,
\begin{equation}
E_{\rm min,max} = \frac{\sqrt{s}}{4} \; \frac{\mse{1}^2-\mneu{1}^2}{\mse{1}^2
} \;
        \left( 1 \pm \sqrt{1- 4 \mse{1}^2/s} \right). \label{eq:edges} \\
\end{equation}
In Ref.~\cite{martyn}, this has been simulated in detail for the Snowmass
parameter point
SPS1a. The expected error for the determination of the spectrum edges can be
extrapolated from that study by scaling the error with the differential
cross-section. This is based on the observation that the total error is
dominated by statistics, which under the assumption of Poisson statistics scales
with the square root of the differential cross-section. The masses can then be
derived from the measured values for the edge locations $E_{\rm min,max}$ using
eq.~\eqref{eq:edges}, leading to
\begin{equation}
\mse{1} = 204.2 \pm 0.09 \gev, \qquad \mneu{1} = 133.4 \pm 0.16 \gev.
\label{eq:massfromspec}
\end{equation}
The extracted values for $\mse{1}$ and $\mneu{1}$ are strongly correlated. Thus
the precision can be further improved by using an independent measurement of
the selectron mass from a $e^-e^-$ threshold scan~\cite{slep,lcws04}.  From the
threshold scan with a total luminosity of 5 fb$^{-1}$, 
a precision of 85 MeV for the selectron mass is expected,
leading in combination with eq.~\eqref{eq:massfromspec} to
\begin{equation}
\mneu{1} = 133.4 \pm 0.13 \gev.
\end{equation}
By including other channels, {\it i.e.} smuon decay spectra, the neutralino mass error
can be further reduced to
\begin{equation}
\mneu{1} = 133.4 \pm 0.11 \gev.
\end{equation}

The most precise method for the determination of the other 
neutralino and chargino masses
at a linear collider is from threshold scans. This has been studied in
experimental simulations for the Snowmass parameter point SPS1a in
Ref.~\cite{lhclc}. A simple estimate of the expected error in the
scenario defined in the appendix can be achieved by rescaling the results of
Ref.~\cite{lhclc} with the different cross-sections near threshold.
Since the mass measurement error is dominated by statistics, the error should
scale with the square root of the cross-section near threshold, assuming Poisson
statistics. It assumed that for each threshold scan an integrated luminosity of
100 fb$^{-1}$ is invested, corresponding to a running time of a few months.

In the SPS1a scenario, the dominant decay mode of the chargino is into a tau
lepton, neutrino and neutralino, $\cha_1^\pm \to \tau^\pm \nu_\tau \neu_1$.
In  the stop co-annihilation scenario, the
chargino decays into the light stop, $\cha^+_1 \to \tilde{t}_1 \, \bar{b}$, is
also open.  For the scenario under study, the branching ratio is  BR$(\cha^+_1
\to \tilde{t}_1 \, \bar{b}) = 98\%$. 

Since the stop decay is thus the dominant
decay mode, the chargino pair production signal is characterized by two b-quark
jets and two c-quark jets and missing energy originating from the neutralinos
in the decay cascade, $e^+e^- \to \cha^+_1 \cha^-_1 \to \tilde{t}_1 \bar{b}
\, \tilde{t}_1^* b \to c \neu_1 \bar{b} \, \bar{c} \neu_1 b$
. While a detailed experimental simulation for this channel
is not available, we estimate the expected signal efficiency using results of
other studies. The signal identification relies heavily on flavor
identification. Demanding a purity of about 80\% each, tagging efficiencies of
about 80\% and 40\% for b-quark and c-quark pairs, respectively, are achievable
\cite{higgspair}. The main backgrounds are triple gauge boson production, which
is small \cite{3gaugebos}, and $t\bar{t}$ events. Both can be reduced by
applying simple cuts on the di-jet invariant masses and the $t\bar{t}$
background is strongly diminished by demanding large missing energy and an
isolated lepton veto. Thus a resulting signal efficiency of 30\% seems
feasible, which is comparable with the efficiency for the decay of the
charginos into taus in the SPS1a scenario \cite{lhclc}. Therefore the
expected precision for the  chargino mass measurements from a threshold scan
can be estimated from the results of Ref.~\cite{lhclc} by simply factoring
in the different cross-section.

The main decay mode for the neutralino $\neu_2$ in the scenario
in the appendix is the leptonic decay $\neu_2 \to l^+l^- \, \neu_1$, $l =
e,\mu,\tau$, as in the SPS1a scenario. Therefore the background reduction
is expected to be similar as described in Ref.~\cite{lhclc} and a
comparable resulting experimental efficiency can be assumed. The main
difference to be taken into account is therefore again the value of the
production cross-section $e^+e^- \to \neu_1 \neu_2$ near threshold.

For the heavier neutralino states, no detailed simulations exist so far.
Preliminary studies \cite{nauenberg} suggest that if the decay channels into gauge
bosons, $\neu_{3,4} \to Z \neu_i, \, W^\pm \cha_1^\mp$, are dominant, the neutralino
mass can be reconstructed from their decay spectra with a precision of 3--5
GeV. Here an error of 4 GeV is assumed for the mass of $\neu_3$, which is
produced in $e^+e^- \to \neu_1 \neu_3$. The neutralino $\neu_4$ and the heavier
chargino $\cha_2^\pm$ are too heavy to be studied at a 500 GeV collider.

Under these assumption the following error estimates are obtained:
\begin{equation}
\delta \mneu{1} = 0.11 \gev, \quad \delta \mneu{2} = 2.5 \gev, \quad
\delta \mneu{3} \approx 4 \gev, \quad \delta \mcha{1} = 0.12 \gev.
\end{equation}
Note that the precision for the chargino mass is substantially better than in
Ref.~\cite{lhclc}, where $\delta \mcha{1} = 0.55$ GeV was obtained. This can be
explained by the fact that for the SPS1a scenario there is a large cancellation
between the s- and t-channel contributions to chargino production. In the
scenario defined in the appendix, on the other hand, the t-channel is suppressed
due to the large sneutrino mass, so that the cancellation does not occur and the
cross-section is about 20 times larger.

The mass measurements are not sufficient to extract the fundamental
supersymmetry parameters without ambiguity.
Additional information is obtained from measurements of the chargino and
neutralino cross-sections.
The chargino cross-section is relatively large, about 1 pb, depending on the
beam polarization, and gives important information through the chargino
couplings and mixings. Using the possibility of beam polarization, two
independent cross-section measurements with left/right- and
right/left-polarized $e^-$/$e^+$ beams can be performed at $\sqrt{s} = 500$
GeV, $\sigma^{\cha_1^+\cha_1^-}_\LL(500)$ and
$\sigma^{\cha_1^+\cha_1^-}_\RR(500)$.

As discussed above, the chargino decays dominantly into the light stop,
$\cha^+_1 \to \tilde{t}_1 \, \bar{b}$. As the branching ratios are difficult
to  determine experimentally, here only the larger chargino decay channel
$\cha^+_1 \to \tilde{t}_1 \, \bar{b}$ will be considered, and for the parameter
extraction only cross-section ratios, not absolute cross-sections, are used.
Besides the ratio of the cross-section for different beam polarizations,
$\sigma^{\cha_1^+\cha_1^-}_\LL(500)/\sigma^{\cha_1^+\cha_1^-}_\RR(500)$, ratios
of the cross-section at other center-of-mass energies do not yield any
significant additional information in the scenario in the appendix. This
can be understood by the fact that the dependence of the cross-section on the
center-of-mass energy is small since the t-channel contributions is switched
off due to the large sneutrino mass.

For the measurement of $\sigma^{\cha_1^+\cha_1^-}_\LL(500)$ and
$\sigma^{\cha_1^+\cha_1^-}_\RR(500)$, a luminosity of 250 fb$^{-1}$ each is assumed. As
before, 80\% $e^-$ and 60\% $e^+$ beam polarization are used.
As discussed above, due to relatively small backgrounds, it is estimated that an
overall signal efficiency of 30\% can be achieved for the chargino decay mode
into stops. Systematic
errors mainly arise from the following sources:  
\begin{itemize}

\item
The error on the chargino mass $\delta
\mcha{1} = 0.12 \gev$ from a threshold scan \cite{lhclc}.

\item The error on the electron-sneutrino mass $m_{\tilde{\nu}_e}$, which enters
into the production process through the t-channel exchange
contribution. Since the sneutrino is heavy, only a lower bound on the mass can
be set, which is assumed to be $m_{\tilde{\nu}_e} > 1000$ GeV.

\item
As above, an uncertainty of $\delta P(e^\pm)/P(e^\pm) = 0.5\%$ is assigned for
the beam polarization~\cite{power}.

\end{itemize}
The cross-section ratio and the two chargino masses are related to the
fundamental parameters $M_2$, $|\mu|$, $\phi_\mu$ and $\tan\beta$ by
eqs.~\eqref{eq:cmass} and the formulae in Ref.~\cite{chath}.

Evidently, three observables are not
sufficient to extract these four parameters. 
In addition, the heavy chargino $\cha^\pm_2$ is difficult to
access at the linear collider, while the identification of charginos at the LHC
is difficult because the hadronic chargino decay is faced with large backgrounds. 
Moreover, the chargino observables exhibit
substantial correlations. For these reasons, and to constrain the missing
parameter $M_1$, also neutralino observables are included in the analysis.

Neutralinos can be produced in various pair combinations, $e^+e^- \to \neu_i
\neu_j$, $i,j = 1 \dots 4$. However, the production rates for $\neu_3$ and
$\neu_4$ are small and the heavy states decay via complex cascades.
Therefore, here only the production cross-sections of the two lighter states are
considered, $e^+e^- \to \neu_1 \neu_2$ and $e^+e^- \to \neu_2 \neu_2$. In
contrast to the charginos, the $\neu_2$ cannot decay into stop quarks, leaving
only the leptonic decay modes open. This allows to measure absolute
cross-sections. For $\sigma(e^+e^- \to \neu_1 \neu_2)$, an experimental
simulation was performed in Ref.~\cite{ball}, yielding an efficiency of 25\%
in SPS1a.
For the process $e^+e^- \to \neu_2 \neu_2$ no detailed simulation exists. It
leads to a final state of four leptons (mainly taus for the case of large
$\tan\beta \gesim 5$) and two LSPs, which result in missing energy in the
detector. This signature has very little background. From the tau tagging
efficiency achieved in Ref.~\cite{ball}, it is expected that  $\sigma(e^+e^- \to
\neu_2 \neu_2)$ can be reconstructed with an efficiency of 15\%.

The two neutralino cross-sections $\sigma(e^+e^- \to \neu_1 \neu_2)$ and
$\sigma(e^+e^- \to \neu_2 \neu_2)$ can be measured 
again for the two polarization combinations $P(e^-)$/$P(e^+)$ = $-$80\%/+60\%
and +80\%/$-$60\% at $\sqrt{s} = 500$ GeV.

In total this amounts to four observables from the neutralino
cross-sections. Systematic errors due to the uncertainty of the selectron
masses in the t-channel contributions and of the beam polarization are taken
into account. The expected precision for the light selectron mass is $\delta
\mse{1} = 0.085 \gev$ from a threshold scan, while for the heavy selectron mass
only a lower bound can be set, which is assumed to be $\mse{2} > 1000$ GeV. For
the polarization uncertainty the same value $\delta P(e^\pm)/P(e^\pm) = 0.5\%$
as above is taken.

\begin{figure}[tb]
\centering
\epsfig{file=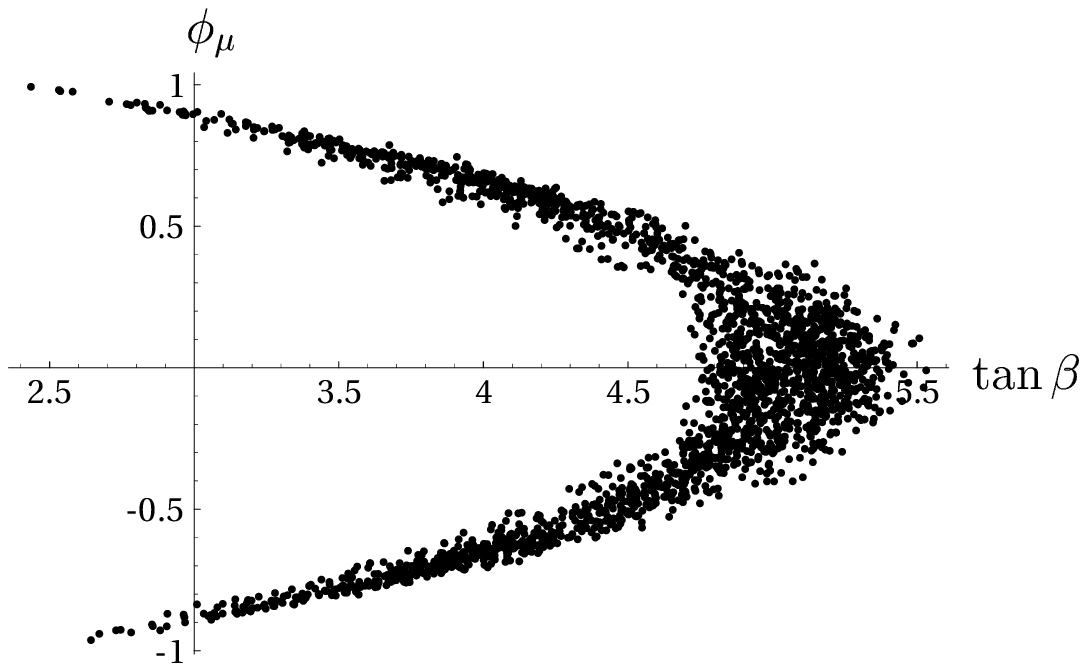, width=9cm}
\mycaption{Allowed region in the $\tan\beta$-$\phi_\mu$ parameter plane from
measurements in the chargino/neutralino sector. 
The black dots correspond to points from a random scan over all parameters that
are allowed by the condition $\Delta \chi^2 \leq
1$.}
\label{fig:corr}
\end{figure}

The neutralino cross-sections and masses can be related to the parameters
$M_1$, $M_2$, $|\mu|$, $\phi_\mu$ and $\tan\beta$ using the formulae of Ref.~\cite{ckmz}.
The chargino and neutralino cross-section and mass measurements are combined in
a $\chi^2$ fit, thus taking into account all correlations. The fit results are
\begin{equation}
\begin{aligned}
M_1 &= (112.6 \pm 0.2) \gev, \\
M_2 &= (225.0 \pm 0.7) \gev,  \\
|\mu| &= (320.0 \pm 3.0) \gev, \\
 |\phi_\mu| &< 1.0, \\
\tan\beta &= 5^{+0.5}_{-2.6}.
\end{aligned}
\end{equation}
Due to correlations, the resulting 1$\sigma$ bounds on the fundamental MSSM
parameter are not very precise for some of the parameters.
In particular, for the given scenario defined in the appendix,  there is a large
correlation between $\tan\beta$ and the CP-violation phase $\phi_\mu$, as shown
in Fig.~\ref{fig:corr}. As a consequence a precise determination of the two
parameters individually is difficult from the chargino and neutralino sector
alone. \label{chaneu}
Ref.~\cite{htb} suggests that $\tan\beta$ can be determined accurately in the
Higgs sector, but this has not been studied for the case of CP violation so far.


\subsection{Higgs parameters}

The s-channel Higgs resonances could be important for the dark matter
annihilation cross-section, hence the Higgs parameters have to determined
independently. In the present scenario give in the appendix, the Higgs and
neutralino masses are such that none of the Higgs states forms a resonance in
the neutralino annihilation.  The experimental verification of this scenario
requires at least a rough measurement of the Higgs masses and couplings.  

By studying the Higgs-strahlung process $e^+ e^- \to Z h^0$, the mass of a
light Higgs boson  with $\MH \approx 120$ GeV can be determined with an error
of less than 100 MeV \cite{higs62,higs54} and the couplings of the Higgs to
Standard Model fermions can be extracted with a precision of a few percent
\cite{higs54,higs53}. The Higgs-neutralino coupling is difficult to measure,
but it can be calculated from the fermionic couplings within the MSSM. In
addition, a model-independent upper bound on the Higgs-neutralino coupling
follows from requiring this coupling to be in the perturbative regime.
This information on the Higgs mass and couplings is sufficient to verify that
the impact of Higgs exchange on the annihilation process is completely
negligible in the given scenario in the appendix, since the annihilation
proceeds far away from the Higgs resonance.

The heavier Higgs states cannot be studied at a 500 GeV linear collider.
However, from searches for the process $e^+e^- \to h^0 A^0$, a lower bound on
the pseudoscalar mass of about $m_{\rm A^0} \sim 380 \gev$ can be established, which
is sufficient to rule out a significant contribution from the heavy Higgs bosons
to the annihilation cross-section.


\section{Dark matter prediction}
\label{sec:dm}

The experimental data from collider experiments discussed in the previous
section can be used to compute the  expected cosmological dark matter relic
density from supersymmetric sources.  The given  sample scenario
in the appendix lies in the region where neutralino-stop co-annihilation is 
effective. Therefore it is expected that a precise prediction will dominantly
rely on the accurate determination of the stop-neutralino mass difference.

While most of the individual measurements are quite precise, large correlations
are found in the extraction of the chargino/neutralino parameters, as discussed
at the end of section \ref{chaneu}. However, since the dark matter computation
mainly depends on the LSP mass $\mneu{1}$, which is measured directly, the
resulting uncertainty due to experimental errors in the chargino/neutralino
sector is relatively small when taking into account the correlations properly.

The relic dark matter density is computed with the program described in
Ref.~\cite{morr} and has been checked against the code ISAReD presented in
Ref.~\cite{isared}. The relevant parameters used as input and derived from the 
experimental analyses are $\mst$, $\cos \theta_{\tilde{t}}$,  $M_1$, $M_2$,
$|\mu|$, $\phi_\mu$, $\tan\beta$, $\sin\alpha$ and $m_{\rm h^0}$. The mass of
the heavier stop $\tilde{t}_2$ is too large to be measured directly, but it is
assumed that a limit of $m_{\tilde{t}_2} > 1000$ GeV can be set from collider
searches. In principle all other MSSM parameters have some effect on the dark
matter annihilation and should be taken into account, but for the given
scenario in the appendix, their contribution is sub-dominant and even rough
estimates or lower mass bounds will be sufficient for an accurate computation
of $\Omega_{\rm CDM}$. Therefore the influence of these other parameters has
been neglected here.

\begin{figure}[tb]
\centering
\epsfig{file=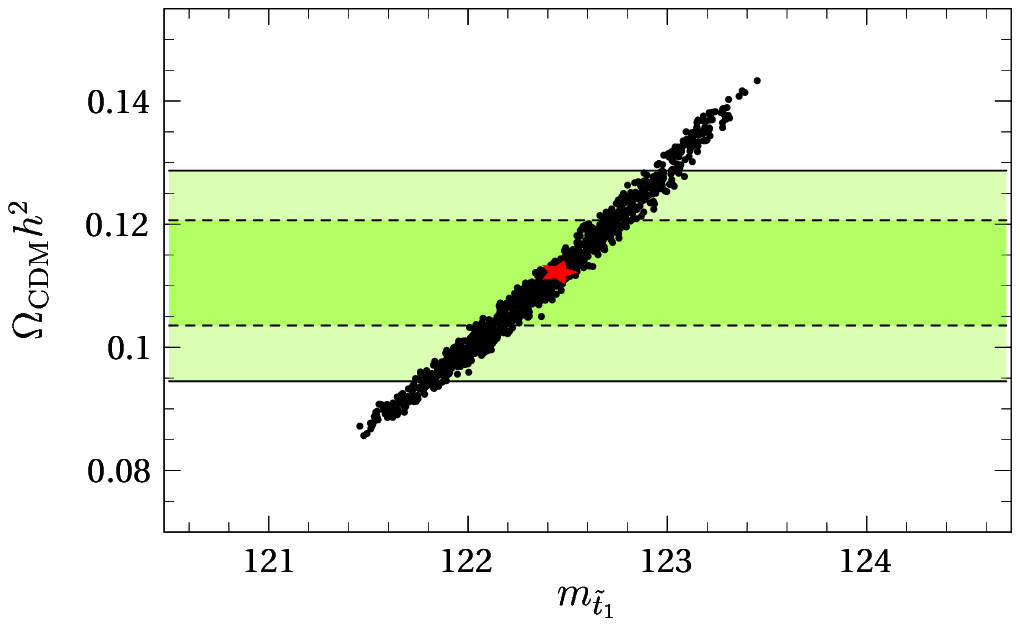, width=13cm}
\mycaption{Computation of dark matter relic abundance $\Omega_{\rm CDM} h^2$
taking into account estimated experimental errors for stop, chargino, neutralino
and Higgs sector measurements at future colliders. The black dots correspond to
a scan over the 1$\sigma$ ($\Delta \chi^2 \leq 1$) region allowed by the 
experimental errors, as a
function of the measured stop mass, with the red star indicating the best-fit
point. The horizontal shaded bands show the
1$\sigma$ and 2$\sigma$ constraints on the relic
density measured by WMAP.}
\label{fig:dm}
\end{figure}
The expected experimental errors are propagated and correlations are taken into
account by means of a $\chi^2$ analysis. No theoretical errors due to higher
order contributions are taken into account in the calculation of the dark
matter density. The result for the scenario given in the appendix is shown in
Fig.~\ref{fig:dm}.  The scattered dots indicate a scan of 100000 random points
in the parameter space allowed by the collider experimental results, as a
function of the measured stop mass. The range of the horizontal axis is
constrained by the error in the stop mass measurement, $m_{\tilde{t}_1} =
(122.5 \pm 1.0) \gev$. The horizontal bands depict the relic density as
measured by WMAP \cite{Spergel:2003cb} with one and two standard deviation
errors. At 1$\sigma$ level, the astrophysical observations lead to $0.104 <
\Omega_{\rm CDM} h^2 < 0.121$.

In total, using the collider measurements of the stop and chargino/neutralino
as input to compute the dark matter annihilation cross-section, the relic
density could be predicted to $0.086 < \Omega_{\rm CDM} h^2 <
0.143$ at the 1$\sigma$ level. Thus the overall precision is of the same
magnitude as, but worse by roughly a factor 3 than
the direct WMAP determination. The uncertainty in the theoretical
determination is dominated by the measurement of the $\tilde{t}_1$ mass.

The precision for the determination of the dark matter density from collider
data could in principle be increased by including other stop mass measurements.
Different methods for the $\tilde{t}_1$ mass measurement have been analyzed in
Ref.~\cite{stopmass}. Besides using cross-section measurements with two beam
polarization combinations, as detailed above, $\mst$ can also be obtained from
a scan near the $\tilde{t}_1 \tilde{t}_1^*$ production threshold or from the
charm jet energy distributions in the decay $\tilde{t}_1 \to c \, \neu_1$. In
preliminary studies of these methods, accuracies of the order of 1--2 GeV were
obtained~\cite{stopmass}, which does not improve on the stop mass measurement
from the total production cross-section.

However, first results of an optimized threshold scan method indicate that the
precision for  $\mst$ can be improved to about 0.5 GeV~\cite{mschmitt}. The
advantage of the threshold scan method is the small influence of systematic
errors, since it makes use of the shape of the cross-section as a function of
the center-of-mass energy, instead of absolute cross-section measurements.
However it is limited by small statistics near the threshold. The best accuracy
can be achieved by combining a measurement near the threshold, where the
cross-section is most sensitive to the stop mass, with a measurement at higher
energies, where the cross-section is larger~\cite{mschmitt}.

With an stop mass error of $\delta \mst = 0.5 \gev$, the relic density could be
computed much more precisely, yielding the result $0.099 < \Omega_{\rm CDM} h^2
< 0.125$ in the scenario defined in the appendix. This precision is very
comparable to the direct WMAP determination.

The sample scenario in the appendix represents just one particular possible
supersymmetry parameter point that has been chosen to give the correct dark
matter relic density. It is also interesting to explore other cases, with
different values of the stop mass and where the dark matter abundance derived
from collider measurements would not overlap with the value obtained
from astrophysical experiments. Table~\ref{tab:points} lists six examples of
supersymmetry scenarios that differ mainly in the stop parameters and the
stop-neutralino mass difference $\Delta m = m_{\tilde{t}_1} - \mneu{1}$. All
points are in agreement with the baryogenesis conditions of a strongly first
order electroweak phase transition, $v(T_{\rm c})/T_{\rm c} \gesim 1$, and a
sufficiently large baryon asymmetry, $\eta \sim 0.6 \times 10^{-10}$. The
constraints from the bounds on the Higgs mass from LEP, $m_{\rm h^0} \gesim
114.4
\gev$, and on the electron electric dipole moment are also taken into account.
\renewcommand{\arraystretch}{1.2}%
\begin{table}[tb]
\centering
\begin{tabular}{|l|cccccc|}
\hline
 &  A & B & C & D & E & F \\
\hline
$m^2_{\rm\tilde{U}_3}$ [GeV$^2$] & 
  $-99^2$ & $-99^2$ & $-99^2$ & $-97^2$ & $-90.5^2$ & $-85.5^2$ 
\\
$m_{\rm\tilde{Q}_3}$ [GeV] &
  2700 & 3700 & 4200 & 4900 & 4700 & 4300
\\
$A_t$ [GeV] &
  $-860$ & $-1150$ & $-1050$ & $-500$ & $-400$ & 0
\\
$M_1$ [GeV] &
  107.15 & 111.6 & 112.6 & 119.0 & 123.2 & 129.0 
\\
$\tan\beta$ &
  5.2 & 4 & 5 & 6 & 5.5 & 5.5 
\\
$A_{e,\mu,\tau}$ &
  $5 \times e^{i \pi/2}$ & $3.7 \times e^{i \pi/2}$ & $5 \times e^{i \pi/2}$ & 
  $5.8 \times e^{i \pi/2}$ & $5.2 \times e^{i \pi/2}$ & $5 \times e^{i \pi/2}$ 
\\
\hline
$m_{\tilde{t}_1}$ [GeV] &
  117.1 & 118.0 & 122.5 & 130.2 & 135.2 & 139.4
\\
$\mneu{1}$ [GeV] &
  102.1 & 104.1 & 107.2 & 114.0 & 118.1 & 123.1
\\
$\cos \theta_{\tilde{t}}$ &
  0.0210 & 0.0150 & 0.0105 & 0.0038 & 0.0035 & 0.0005
\\
$m_{\rm h^0}$ [GeV] &
  115.1 & 115.0 & 117.0 & 117.1 & 116.2 & 115.1
\\
$\Omega_{\rm CDM} h^2$ &
  0.113 & 0.060 & 0.112 & 0.144 & 0.166 & 0.112 
\\
\hline
\end{tabular}
\mycaption{A few exemplary supersymmetry scenarios that are compatible with
electroweak baryogenesis and bounds on the Higgs mass and the electron electric
dipole moment. All other parameters not in the table are as in
eq.~\eqref{eq:scen} in the appendix. The scenario C is identical with the scenario 
in the appendix.}
\label{tab:points}
\end{table}
\renewcommand{\arraystretch}{1}%

The points A, C and F give a dark matter density in good agreement with
current astrophysical observations. Here the scenario C is identical with the
one introduced in the appendix. The point D is marginally consistent
with the WMAP constraints, while the points B and E result in a dark matter
density below and above, respectively, the WMAP measurement by more than the
95\% confidence level.

\begin{figure}[tb]
\centering
\epsfig{file=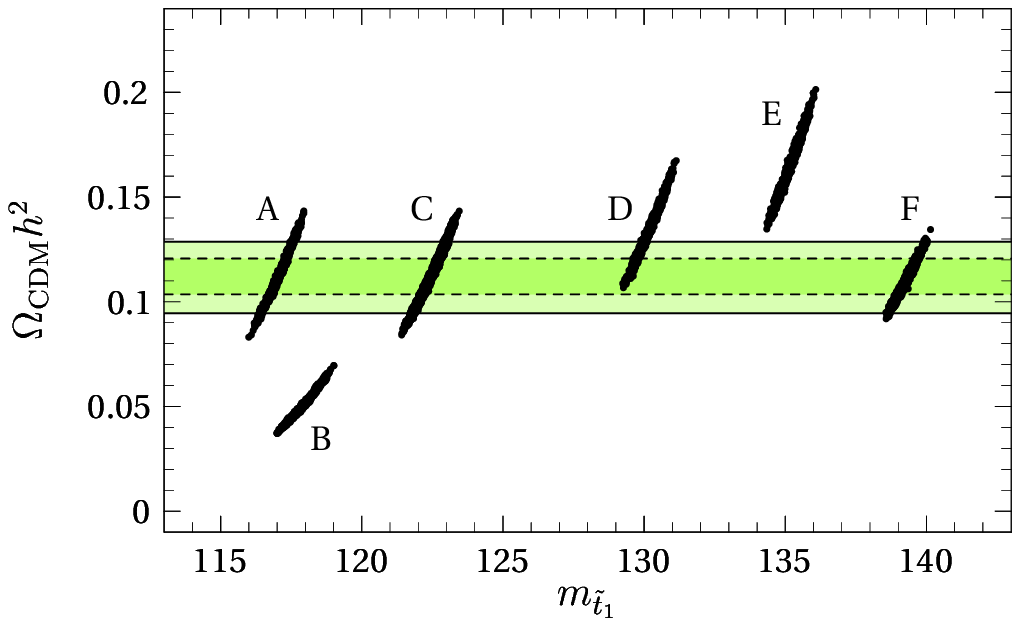, width=13cm}
\mycaption{Supersymmetric dark matter relic abundance $\Omega_{\rm CDM} h^2$
predicted from collider stop, chargino, neutralino and Higgs sector
measurements in different assumed supersymmetry scenarios from
Table~\ref{tab:points}. The black dots correspond $\chi^2$ scans over the
supersymmetry parameter region  allowed by projected 1$\sigma$ experimental
errors. The horizontal shaded bands show the 1$\sigma$ and 2$\sigma$
constraints on the relic density measured by WMAP.}
\label{fig:dm2}
\end{figure}

Performing the analysis of experimental uncertainties as explained in the
previous section for each of the points in Table~\ref{tab:points} leads to the 
results shown in Fig~\ref{fig:dm2}.
As evident from the figure, the linear collider measurements constrain the
computed dark matter relic density with a precision comparable to the
current direct astrophysical observation. For the points A, C and F this would
indicate strong evidence that supersymmetry with a LSP neutralino and a light
stop contributing to the co-annihilation mechanism is the source of 
dark matter in the universe. 

In case of scenario D, the linear collider data would restrict the dark matter
abundance to $0.107 < \Omega_{\rm CDM} h^2 < 0.167$ within 1$\sigma$
experimental errors. While this result would still be consistent with the
current astrophysical result $0.095 < \Omega_{\rm CDM} h^2 < 0.129$ from WMAP,
it imposes constraints  on the supersymmetric parameter space. Under the
assumption that our understanding of the cosmological evolution is correct, the
$\tilde{t}_1$ mass is for example required to be less than about 130 GeV.

For point E, the deduced neutralino dark matter density turns out to be too
large compared to the WMAP result by roughly two standard deviations. This
could be due to other particles contributing to increase the total dark matter
annihilation cross-section. It can also be interpreted as evidence that our
current theoretical understanding of the evolution of the universe needed to be
revised. On the other hand, scenario B leads to a dark matter density that is
smaller than the WMAP result by about two standard deviations. This discrepancy
could be explained by the existence of another source for cold dark matter or,
as before, might put our description of cosmological evolution into question.


\section{Conclusions and outlook}
\label{sec:concl}

The MSSM with light scalar top quarks provides a framework for simultaneously
explaining electroweak baryogenesis and the correct  dark matter relic density.
The  stop-neutralino co-annihilation mechanism can play an important part in
this picture. The exploration of this
scenario relies on precision measurements at accelerator experiments and will
be within reach for the next generation of colliders, in particular the ILC.

Based on a detailed experimental simulation including detector effects, it was
shown that light stop quarks can be discovered at the ILC for stop masses up to
almost the beam energy and for stop-neutralino mass differences $\Delta m =
\mst - \mneu{1}$ down to about 5 GeV. In particular, the complete parameter
region where stop-neutralino co-annihilation is effective can be covered. Since
stops with small $\Delta m$ are difficult to explore at hadron colliders, the
ILC thus provides a unique discovery capability for this kind of scenario.

In addition, it was found that the stop mass and mixing angle can be accurately
determined from measurements of the stop production cross-section for two
different beam polarizations at the ILC. Together with precision measurements
in the chargino and neutralino sector,  this would allow to compute the dark
matter relic density from experimental results at the ILC in a bottom-up
approach, {\it i.e.} without assuming a specific mechanism or pattern for
supersymmetry breaking parameters. Including statistical and systematic errors
in the experimental analysis,  it was estimated that the precision for the
relic density induced from ILC measurements is slightly worse, but comparable
with the current direct determination from WMAP and SDSS.
Refinements in the determination of the stop mass can improve this
result significantly.

In MSSM scenarios that are consistent with electroweak baryogenesis, the lower
limit on the Higgs boson mass from searches at LEP and bounds from electric
dipole moments, the sfermions of the first two generations need to be
relatively heavy. As a consequence, for these scenarios the exploration of
supersymmetric partners and their properties is difficult at the LHC.
Nevertheless, the LHC can establish lower limits on the masses of heavy
superpartners, thus providing important information for the understanding and
computation of the dark matter density in the universe.

The present scenario with light scalar top quark and  small stop-neutralino
mass differences is also a motivation for refining the development of the
vertex detector. The analysis is based on a fast and realistic simulation of a
detector which includes a CCD vertex tracker. The importance of the vertex
detector performance for c-quark tagging in scalar top quark decays has been 
discussed in the framework of the Linear Collider Flavour Identification (LCFI)
collaboration which studies CCD detectors for quark flavour identification.
Regarding the vertex detector design, the study of c-quark tagging with small
visible energy, as expected from the scalar top decays in the discussed
scenario, is particularly challenging and could serve well as a benchmark
reaction.

This paper is the first step in a larger project aiming to explore in detail
the collider signatures and measurements that can elucidate the origin of
baryonic and dark matter in the universe within the framework of supersymmetry.
Future avenues include refinements in the experimental analysis, especially for
small mass differences $\Delta m$, and the investigation of the impact of the
vertex detector layout on the charm tagging performance.

In this work, the experimental analysis was designed with a selection procedure
that is independent of the actual values of the unknown supersymmetry
parameters. Such a search strategy, which does not require any prior knowledge
of the supersymmetry scenario, is important for a model-independent discovery
potential. However, once evidence for a stop signal would be found and the stop
and neutralino mass would be measured with some precision, the analysis can be
refined using this knowledge, in order to achieve a better separation of signal
and background and increase the precision of the stop parameter determination.

In the present analysis, the projected uncertainty in the computation of the
dark matter abundance is dominated by the error in the stop mass. By including
other methods for the stop mass measurement, the precision for the dark matter
computation could possibly be improved substantially. A promising possibility
is a method making use of cross-section measurements near the pair production
threshold, which seems to be able to reduce the stop mass error by a factor of
two.

The present work has been performed using tree-level formulae and
cross-sections for the ILC analysis and the computation of the dark matter
annihilation rate. For the expected experimental precision, however, radiative
corrections are important and introduce a dependence on other supersymmetry
parameters, {\it e.g.} sfermion masses and mixing outside of the stop sector.
This issue will be studied in future work.

In this publication a specific MSSM scenario was studied in detail for the
experimental determination of parameters and the dark matter analysis. In
future work, this analysis shall be extended to other scenarios that are
similarly cosmologically motivated, but have different collider signatures.

The study presented in this report opens a window toward exploring some of
the main unsolved questions in the evolution of the universe by establishing
cross-relations between collider measurements and cosmological processes and by
combining experimental analyses and theoretical computations. A sophisticated
phenomenological program with precision measurements of new physics properties
at the ILC sets the base for deepening our understanding of the cosmos.


\appendix

\section*{Appendix: MSSM case study scenario} 
\label{sec:app}

The numerical analysis in sections \ref{sec:param} and \ref{sec:dm} are based on
a specific MSSM parameter point characterized by the following values:
\begin{equation}
\begin{aligned}
m^2_{\rm\tilde{U}_3} &= -99^2 \gev^2,	& M_1 &= 112.6 \gev, \\
m_{\rm\tilde{Q}_3} &= 4200 \gev,	& M_2 &= 225 \gev, \\
m_{\rm\tilde{D}_3} &= 4000 \gev,	& |\mu| &= 320 \gev,\\
A_t &= -1050 \gev,		& \phi_\mu &= 0.2, \\
m_{\rm\tilde{L}_{1,2,3}} &= 2000 \gev,	& \tan\beta &= 5, \\
m_{\rm\tilde{R}_{1,2,3}} &= 200 \gev,	& m_{\rm A^0} &= 800 \gev,\\
A_{e,\mu,\tau} &= 5 \tev \times e^{i \pi/2}, \\
m_{\rm\tilde{Q},\tilde{U},\tilde{D}_{1,2}} &= 4000 \gev.
\end{aligned}
\label{eq:scen}
\end{equation}
The potentially large contribution of one-loop sfermion-neutralino and
sfermion-chargino loops to the electric dipole moments of the electron and
neutron requires the  sleptons and squarks of the first two generations to be
heavy. Therefore all squark soft supersymmetry breaking parameters of the first
two generations are taken to be 4 TeV, and in addition, the left-chiral
sleptons are also assumed to be heavy. The right-chiral sleptons contribute to
the electron electric dipole moment on a sub-leading level and can be taken
with a mass of a few hundred GeV if the selectron $A$-parameter carries a
CP-violating phase that cancels part of the effect of the phase of $\mu$. Here,
the right-chiral slepton soft supersymmetry breaking mass is assumed to be 200
GeV, resulting in physical R-slepton masses within the kinematical reach of a
500 GeV linear collider.

The chosen parameters are compatible with a strongly first order electroweak
phase transition for electroweak baryogenesis, $v(T_{\rm c})/T_{\rm c} \gesim
1$~\cite{CQW}, generate a sufficiently large baryon asymmetry, $\eta \sim 0.6
\times 10^{-10}$, and yield a value for the dark matter relic
abundance\footnote{The relic dark matter density has been computed with the
code used in Ref.~\cite{morr}.} within the WMAP bounds, $\Omega_{\rm CDM} h^2 =
0.1122$. In addition, the supersymmetry parameters, in particular the stop
parameters, are chosen such that the mass of the lightest Higgs boson is $m_{\rm
h^0} = 117$ GeV,  sufficiently above the
bound from direct searches at LEP $m_{\rm h^0} \gesim 114.4 \gev$
\cite{lephbound}. At tree-level the following masses are obtained for the
relevant supersymmetric particles:
\begin{equation}
\begin{aligned}
m_{\tilde{t}_1} &= 122.5  \gev, & m_{\tilde{e}_1} &= 204.2 \gev, &
\mneu{1} &= 107.2 \gev, & \mcha{1} &= 194.3 \gev, \\
m_{\tilde{t}_2} &= 4203 \gev,	& m_{\tilde{e}_2} &= 2 \tev, &
\mneu{2} &= 196.1 \gev,	& \mcha{2} &= 358.1 \gev, \\
\cos \theta_{\tilde{t}} &= 0.0105, & m_{\tilde{\nu}_e} &= 2 \tev, &
\mneu{3} &= 325.0 \gev,	\\
& & & &
\mneu{4} &= 359.3 \gev.
\end{aligned}
\end{equation}


\section*{Acknowledgments}

The authors greatly benefited from practical advice by S.~Mrenna about {\sc
Pythia}, by P.~Bechtle about {\sc Simdet} and by T.~Kuhl about the c-tagging
procedure. Special thanks go to C.~Bal\'azs for useful discussions and
assistance with the program ISAReD, and to M.~Schmitt for proofreading
of the manuscript and very valuable comments.


\end{document}